\pdfoutput=1
\documentclass[12pt,fleqn]{article}

\usepackage{ifthen}
\newboolean{FULL}
\setboolean{FULL}{true}

\usepackage{comment}
\usepackage[letterpaper]{geometry}

\usepackage{amsmath,amsthm,mathtools,thm-restate}
\usepackage{fullpage}
\usepackage{authblk}

\usepackage[T1]{fontenc}
\usepackage{esvect}  % \vv
\usepackage{stmaryrd}  % \llbracket \rrbracket
\usepackage[Symbolsmallscale]{upgreek}  % \uppsi, etc.
\usepackage{mathrsfs}  % \mathscr

\usepackage{amssymb,amsfonts}

\usepackage[style=alphabetic,natbib=true,sortcites=true,backref=true,maxnames=99,maxalphanames=99]{biblatex}

\DefineBibliographyStrings{english}{%
  backrefpage = {$\Lsh$~p.},% originally "cited on page"
  backrefpages = {$\Lsh$~pp.},% originally "cited on pages"
}
\usepackage[colorlinks=true,citecolor=blue!65!black,linkcolor=red!75!black]{hyperref}
\usepackage[nameinlink]{cleveref}

\usepackage{booktabs,colortbl}
\usepackage{diagbox,enumitem}
\usepackage{framed,fancybox,ascmac}
\usepackage[normalem]{ulem}

\usepackage{comment}
\usepackage{url}

\usepackage[colorinlistoftodos,prependcaption]{todonotes}

\usepackage{xspace}
\usepackage{bm}

\usepackage{tikz}
\usetikzlibrary{positioning}
\usetikzlibrary{decorations.markings}
\usetikzlibrary{shapes.geometric}
\usetikzlibrary{arrows.meta}
\usetikzlibrary{arrows,automata,decorations.pathmorphing,fit,positioning,arrows.meta,calc}

\crefname{theorem}{Theorem}{Theorems}
\crefname{proposition}{Proposition}{Propositions}
\crefname{lemma}{Lemma}{Lemmas}
\crefname{claim}{Claim}{Claims}
\crefname{corollary}{Corollary}{Corollaries}
\crefname{remark}{Remark}{Remarks}
\crefname{observation}{Observation}{Observations}
\crefname{hypothesis}{Hypothesis}{Hypotheses}
\crefname{definition}{Definition}{Definitions}
\crefname{problem}{Problem}{Problems}
\crefname{example}{Example}{Examples}

\crefname{appendix}{Appendix}{Appendices}
\crefname{section}{Section}{Sections}
\crefname{equation}{Eq.}{Eqs.}
\crefname{figure}{Figure}{Figures}
\crefname{table}{Table}{Tables}
\crefname{algorithm}{Algorithm}{Algorithms}

\usepackage[ruled]{algorithm}
\usepackage{algpseudocodex}

\algrenewcommand\textproc{\textsl}

\renewcommand{\geq}{\geqslant}
\renewcommand{\leq}{\leqslant}
\renewcommand{\phi}{\varphi}
\renewcommand{\epsilon}{\varepsilon}

\renewcommand{\tilde}{\widetilde}

\newcommand{\prb}[1]{\textup{\textsc{#1}}\xspace}
\renewcommand{\prb}[1]{\textup{\textsf{#1}}\xspace}
\newcommand{\nth}[1]{#1\textsuperscript{th}\xspace}

\renewcommand{\vec}[1]{\mathbf{\bm{#1}}}

\newcommand{\reco}{\leftrightsquigarrow}
\newcommand{\rme}{\mathrm{e}}

\DeclareMathOperator*{\argmin}{argmin}

\let\Pr\relax\DeclareMathOperator*{\Pr}{\mathbb{Pr}}
\DeclareMathOperator{\bigO}{\mathcal{O}}
\DeclareMathOperator{\val}{\mathsf{val}}
\DeclareMathOperator{\Had}{\mathsf{Had}}
\DeclareMathOperator{\enc}{\mathsf{enc}}

\newcommand{\ini}{\mathsf{ini}}
\newcommand{\tar}{\mathsf{tar}}
\newcommand{\ttt}{t}
\newcommand{\TTT}{T}

\newcommand{\seqsigma}{\vec{\upsigma}}
\newcommand{\seqpsi}{\vec{\uppsi}}

\newcommand{\YES}{\textup{\textsc{yes}}\xspace}
\newcommand{\NO}{\textup{\textsc{no}}\xspace}

\newcommand{\NP}{\textup{\textbf{NP}}\xspace}
\newcommand{\PSPACE}{\textup{\textbf{PSPACE}}\xspace}

\newcommand{\calA}{\mathcal{A}}

\newcommand{\calP}{\mathcal{P}}

\newcommand{\bbF}{\mathbb{F}}

\newcommand{\bbN}{\mathbb{N}}

\newcommand{\scrC}{\mathscr{C}}

\newcommand{\scrS}{\mathscr{S}}

\let\Pr\relax\DeclareMathOperator*{\Pr}{\mathbb{P}}
\DeclareRobustCommand{\lipicsEnd}{%
	\leavevmode\unskip\penalty9999 \hbox{}\nobreak\hfill
	\quad\hbox{$\lrcorner$}%
}

\newtheorem{theorem}{Theorem}[section]
\newtheorem{proposition}[theorem]{Proposition}
\newtheorem{lemma}[theorem]{Lemma}
\newtheorem{claim}[theorem]{Claim}
\newtheorem{corollary}[theorem]{Corollary}
\newtheorem{observation}[theorem]{Observation}

\theoremstyle{definition}
\newtheorem{definition}[theorem]{Definition}
\newtheorem{problem}[theorem]{Problem}
\newtheorem{example}[theorem]{Example}

\newenvironment{claim*}{\begin{claim}}{\end{claim}}

\numberwithin{equation}{section}

\title{Alphabet Reduction for Reconfiguration Problems}
\author{Naoto Ohsaka\thanks{
    CyberAgent, Inc., Tokyo, Japan.
    \href{mailto:ohsaka\_naoto@cyberagent.co.jp}{\texttt{ohsaka\_naoto@cyberagent.co.jp}}; \href{mailto:naoto.ohsaka@gmail.com}{\texttt{naoto.ohsaka@gmail.com}}
}}
\date{\today}

\begin{document}
\maketitle

\thispagestyle{empty}
\begin{abstract}\noindent We present a reconfiguration analogue of \emph{alphabet reduction}
\`{a}~la {Dinur} (J.~ACM, 2007) \cite{dinur2007pcp} and its applications.
Given a binary constraint graph $G$ and its two satisfying assignments $\psi^\mathsf{ini}$ and $\psi^\mathsf{tar}$,
the \prb{Maxmin Binary CSP Reconfiguration} problem requests
to transform $\psi^\mathsf{ini}$ into $\psi^\mathsf{tar}$ by 
repeatedly changing the value of a single vertex so that
the minimum fraction of satisfied edges is maximized.
We demonstrate a polynomial-time reduction from
\prb{Maxmin Binary CSP Reconfiguration} with arbitrarily large alphabet size $W \in \mathbb{N}$
to itself with universal alphabet size $W_0 \in \mathbb{N}$ such that
\begin{enumerate}
\item the perfect completeness is preserved, and
\item if any reconfiguration for the former violates $\varepsilon$-fraction of edges,
then $\Omega(\varepsilon)$-fraction of edges must be unsatisfied during any reconfiguration for the latter.
\end{enumerate}
The crux of its construction is the \emph{reconfigurability of Hadamard codes},
which enables to reconfigure between a pair of codewords, while avoiding 
getting too close to the other codewords.
Combining this alphabet reduction with gap amplification due to {Ohsaka} (SODA 2024) \cite{ohsaka2024gap},
we are able to amplify the $1$~vs.~$1-\varepsilon$ gap for arbitrarily small $\varepsilon \in (0,1)$
up to the $1$~vs.~$1-\varepsilon_0$ for some universal $\varepsilon_0 \in (0,1)$
\emph{without} blowing up the alphabet size.
In particular,
a $1$~vs.~$1-\varepsilon_0$ gap version of \prb{Maxmin Binary CSP Reconfiguration} with alphabet size $W_0$
is \PSPACE-hard
only assuming the Reconfiguration Inapproximability Hypothesis
posed by {Ohsaka} (STACS 2023)~\cite{ohsaka2023gap},
whose gap parameter can be arbitrarily small.
As an immediate corollary,
we show that under the same hypothesis,
there exists a universal constant $\varepsilon_0 \in (0,1)$ such that
many popular reconfiguration problems are \PSPACE-hard to approximate within a factor of $1-\varepsilon_0$,
including those of
\prb{$3$-SAT}, \prb{Independent Set}, \prb{Vertex Cover}, \prb{Clique}, \prb{Dominating Set}, and \prb{Set Cover}.
This may not be achieved only by gap amplification of \cite{ohsaka2024gap},
which makes the alphabet size gigantic depending on the gap value of the hypothesis.
\end{abstract}
\clearpage
%\tableofcontents\clearpage
\section{Introduction}
\subsection{Background}
\label{subsec:intro:background}

\emph{Combinatorial reconfiguration} is a brand-new field
in theoretical computer science that concerns
the reachability and connectivity over
the solution space of a combinatorial problem.
One canonical \PSPACE-complete reconfiguration problem is \prb{Binary CSP Reconfiguration}:
given a binary constraint graph $G$ over alphabet $\Sigma$ and
its two satisfying assignments $\psi^\ini$ and $\psi^\tar$,
we are requested
to transform $\psi^\ini$ into $\psi^\tar$ by
repeatedly changing the value of a single vertex
while the feasibility of intermediate assignments is maintained.
Such a sequence of feasible solutions is referred to as a \emph{reconfiguration sequence}.
Since the establishment of the unified framework due to
\citet{ito2011complexity},
the complexity of many reconfiguration problems
has been investigated, including those of 
\prb{Satisfiability},
\prb{Coloring},
\prb{Independent Set},
\prb{Vertex Cover}, and
\prb{Clique}.
We refer the readers to the survey by
\citet{nishimura2018introduction,heuvel13complexity,bousquet2022survey,mynhardt2019reconfiguration}.
One latest trend is to study
\emph{approximate reconfigurability} \cite{ohsaka2023gap,ohsaka2023approximate,ohsaka2024gap},
which affords to relax the feasibility of intermediate solutions
during reconfiguration.
For example,
in \prb{Maxmin Binary CSP Reconfiguration} \cite{ohsaka2023gap},
which is an \emph{optimization version} of \prb{Binary CSP Reconfiguration},
we can adopt any \emph{non-satisfying assignments}, but
are required to maximize the minimum fraction of edges satisfied during reconfiguration.
Such optimization versions would be come up with naturally to deal with \PSPACE-hardness of many reconfiguration problems.
See \cref{subsec:intro:related} for other optimization versions of reconfiguration problems.

One of the most important questions concerning approximate reconfigurability is
\emph{\PSPACE-hardness of approximation} for reconfiguration problems,
posed by \citet[Section~5]{ito2011complexity} as an open problem.
Though \NP-hardness of approximation for reconfiguration problems (e.g., \prb{Maxmin SAT Reconfiguration}) was shown by 
\cite{ito2011complexity}, their proofs do not imply \PSPACE-hardness
because of relying on the \NP-hardness of approximating the corresponding optimization problems (e.g., \prb{Max SAT}).
The significance of showing \PSPACE-hardness compared to \NP-hardness is that
it disproves the existence of a witness (especially a reconfiguration sequence) of polynomial length under
\NP~$\neq$~\PSPACE.
\citet{ohsaka2023gap} showed that
a host of (optimization versions of) reconfiguration problems are \PSPACE-hard to approximate
under the \emph{Reconfiguration Inapproximability Hypothesis} (RIH),
which postulates that a gap version of \prb{Maxmin CSP Reconfiguration} is \PSPACE-hard.\footnote{
Very recently, \citet{karthik2023inapproximability,hirahara2024probabilistically} independently announced the proof of RIH.
See \cref{subsec:intro:related} for the positioning of the present work given  \cite{karthik2023inapproximability,hirahara2024probabilistically}.}
The present study delves deeper into
\PSPACE-hardness of approximation for reconfiguration problems \emph{assuming} RIH.

The limitation of \cite{ohsaka2023gap}
is that the degree of inapproximability is not explicitly shown:
although RIH implies that
\prb{Maxmin Binary CSP Reconfiguration} is \PSPACE-hard to approximate within a factor of $1-\epsilon$,
RIH itself does not specify any value of the \emph{gap parameter} $\epsilon \in (0,1)$,
which can be arbitrarily small.
To circumvent this limitation, \citet{ohsaka2024gap} successfully developed
\citeauthor{dinur2007pcp}'s style \emph{gap amplification} \cite{dinur2007pcp} for \prb{Maxmin Binary CSP Reconfiguration},
which amplifies
the $1$ vs.~$1-\epsilon$ gap for arbitrarily small $\epsilon \in (0,1)$
up to the $1$ vs.~$0.9942$ gap.
This result can be used to show
$1.0029$-inapproximability for \prb{Minmax Set Cover Reconfiguration} \cite{ohsaka2024gap} under RIH.
Unfortunately, there still remains another issue:
\emph{the alphabet size becomes gigantic depending on the gap parameter $\epsilon$}.\footnote{
Precisely, the alphabet size becomes $W^{d^{\bigO(\epsilon^{-1})}}$ for some $W,d \in \bbN$ by \cite{ohsaka2024gap}, which is doubly exponential in $\epsilon^{-1}$.
}
Consider for example reducing
\prb{Maxmin Binary CSP Reconfiguration} with alphabet size $W$ to
\prb{Maxmin $3$-SAT Reconfiguration}
in a gap-preserving manner.
According to \cite{ohsaka2023gap},
if the former problem has a $\delta$-gap,
the latter problem's gap turns out to be $\frac{\delta}{2^{\Omega(W)}}$.
This is undesirable if $W$ depends on $\epsilon$.
Our target in this paper is thus
a reconfiguration analogue of \emph{alphabet reduction}, i.e.,
a polynomial-time reduction from \prb{Maxmin Binary CSP Reconfiguration} to itself
that makes a large alphabet into a tiny one
without much deteriorating the gap value.

\subsection{Our Results}
\label{subsec:intro:results}

We present alphabet reduction for \prb{Maxmin Binary CSP Reconfiguration}
\`{a}~la \citet{dinur2007pcp} and its applications.
Given an instance of \prb{Maxmin Binary CSP Reconfiguration}
with arbitrarily large alphabet,
we are able to reduce the alphabet size to a universal constant $W_0 \in \bbN$
preserving the gap value by up to a constant factor:

\begin{theorem}[Alphabet reduction; informal; see \cref{thm:ABC-reduct}]
There exist universal constants $W_0 \in \bbN$ and $\kappa \in (0,1)$ and
a polynomial-time reduction from
\prb{Maxmin Binary CSP Reconfiguration} with arbitrarily large alphabet size $W \in \bbN$
to itself with alphabet size $W_0$ such that
\begin{enumerate}
    \item the perfect completeness is preserved, and
    \item if any reconfiguration for the former violates $\epsilon$-fraction of edges,
    then $\kappa \cdot \epsilon$-fraction of edges
    must be unsatisfied during any reconfiguration for the latter.
\end{enumerate}
\end{theorem}\noindent
Our reduction is independent of $\epsilon$; namely,
$\epsilon$ does not have to be constant, e.g., $\epsilon = (\text{\# of edges})^{-1}$.
The main ingredient of its construction is
the \emph{reconfigurability of Hadamard codes}, which appears later in \cref{subsec:intro:overview}.

As a corollary of \cref{thm:ABC-reduct} and \cite{ohsaka2023gap,ohsaka2024gap},
we are able to amplify the $1$ vs.~$1-\epsilon$ gap
for arbitrarily small $\epsilon \in (0,1)$ up to
the $1$ vs.~$1-\epsilon_0$ gap for some universal $\epsilon_0 \in (0,1)$
\emph{without} blowing up the alphabet size.
Slightly more formally,
for any $\epsilon \in (0,1)$ and $W \in \bbN$,
\prb{Gap$_{1,1-\epsilon}$ Binary CSP$_W$ Reconfiguration}
requests to distinguish whether,
for a binary constraint graph with alphabet size $W$ and its two satisfying assignments $\psi^\ini$ and $\psi^\tar$,
(1) there exists a reconfiguration sequence from $\psi^\ini$ to $\psi^\tar$ consisting only of satisfying assignments, or
(2) every reconfiguration sequence violates more than $\epsilon$-fraction of edges.

\begin{table}[t]
    \centering
    \small
    \setlength{\tabcolsep}{3pt}
    \begin{tabular}{ccc|ccc}
        \toprule
        gap problem & ref. & technique & gap value & alphabet size \\
        \midrule
        \prb{$q$-CSP Reconf} \footnotesize{for any $q$}
            & ---  & --- & arbitrarily small $\epsilon$ & arbitrarily large $W$ \\
        \prb{Binary CSP Reconf} & \cite{ohsaka2023gap} &
            degree reduction & depends on $\epsilon,q,W$ & universal const. \\
        \prb{Binary CSP Reconf} & \cite{ohsaka2024gap} &
            gap amplification & universal const. & depends on $\epsilon,q,W$ \\
        \prb{Binary CSP Reconf} & (this paper) &
            alphabet reduction & universal const.~$\epsilon_0$ & universal const.~$W_0$ \\
        \bottomrule
    \end{tabular}
    \caption{
        Gap-preserving reductions used in \cref{cor:ABC-reduct}.
        We can reduce
        \prb{Gap$_{1,1-\epsilon}$ $q$-CSP$_{W}$ Reconfiguration} (i.e., RIH) to
        \prb{Gap$_{1,1-\epsilon_0}$ Binary CSP$_{W_0}$ Reconfiguration}
        regardless of the values of $\epsilon \in (0,1)$ and $q,W \in \bbN$.
    }
    \label{tab:cor:ABC-reduct}
\end{table}

\begin{corollary}[from \cref{thm:ABC-reduct} and \cite{ohsaka2023gap,ohsaka2024gap}]
\label{cor:ABC-reduct}
There exist universal constants
$\epsilon_0 \in (0,1)$ and
$W_0 \in \bbN$ such that
for arbitrarily small $\epsilon \in (0,1)$ and large $q,W \in \bbN$,
there exists a gap-preserving reduction from
\prb{Gap$_{1,1-\epsilon}$ $q$-CSP$_W$ Reconfiguration}
to
\prb{Gap$_{1,1-\epsilon_0}$ Binary CSP$_{W_0}$ Reconfiguration}.
In particular, the latter problem is \PSPACE-hard under RIH.
\end{corollary}\noindent
Since \emph{both} $\epsilon_0$ and $W_0$ do not depend on 
either $\epsilon$, $q$, or $W$,
\cref{cor:ABC-reduct} makes
the degree of inapproximability and alphabet size of \prb{Maxmin Binary CSP Reconfiguration}
\emph{oblivious to} the (arbitrarily small) gap parameter of RIH.
(Concretely, we would have
$\epsilon_0 = \kappa \cdot (1-0.9942) > 10^{-18}$ and $W_0 < 2{,}000{,}000$, where
number $0.9942$ comes from \cite{ohsaka2024gap}.
See also the proof of \cref{thm:ABC-reduct}.)
This may not be achieved only by gap amplification due to \citet{ohsaka2024gap}.
See also \cref{tab:cor:ABC-reduct} for a sequence of gap-preserving reductions 
used in \cref{cor:ABC-reduct}.

By \cref{cor:ABC-reduct},
we immediately obtain the following gap-preserving reducibility from RIH to many popular reconfiguration problems:

\begin{theorem}[from \cref{cor:ABC-reduct} and
\cite{ohsaka2023gap,ohsaka2024gap}]
\label{cor:inapprox-reconf}
There exists a universal constant $\epsilon_0 \in (0,1)$ such that
for every $\epsilon \in (0,1)$ and $q,W \in \bbN$,
\prb{Gap$_{1,1-\epsilon}$ $q$-CSP$_W$ Reconfiguration}
is polynomial-time reducible to
a $1$ vs.~$1-\epsilon_0$ gap version of
the following reconfiguration problems\textup{:}

\begin{center}
\prb{Binary CSP Reconfiguration},
\prb{$3$-SAT Reconfiguration},
\prb{Independent Set Reconfiguration},
\prb{Vertex Cover Reconfiguration},
\prb{Clique Reconfiguration},
\prb{Dominating Set Reconfiguration},
\prb{Set Cover Reconfiguration}, and
\prb{Nondeterministic Constraint Logic}.
\end{center}
In particular,
optimization versions of the above problems are \PSPACE-hard to approximate within a factor of $1-\epsilon_0$
under RIH.
\end{theorem}\noindent
Once again, \cref{cor:inapprox-reconf} is different from
a consequence of gap-preserving reductions from RIH due to \citet{ohsaka2023gap}
in a sense that it renders
$\epsilon_0$ \emph{independent of} the value of $\epsilon$.\footnote{
We stress that \cref{cor:inapprox-reconf} is essentially different from
the following statement
(where $\epsilon_0$ can depend on $\epsilon$, $q$, and $W$),
which is immediate from \cite{ohsaka2023gap}:
``For arbitrarily small $\epsilon \in (0,1)$ and large $q,W \in \bbN$,
there exists $\epsilon_0 \in (0,1)$ such that
\prb{Gap$_{1,1-\epsilon}$ $q$-CSP$_W$ Reconfiguration}
is polynomial-time reducible to
a $1$ vs.~$1-\epsilon_0$ gap version of
the reconfiguration problems listed in \cref{cor:inapprox-reconf}.''
}
Such results (in terms of \PSPACE-hardness under RIH) seem to be known only for (optimization versions of)
\prb{Binary CSP Reconfiguration} ($0.9942$-factor) \cite{ohsaka2024gap},
\prb{Set Cover Reconfiguration} ($1.0029$-factor) \cite{ohsaka2024gap}, and
\prb{Clique Reconfiguration} ($n^{-\epsilon}$-factor) \cite{hirahara2024probabilistically}
(to the best of our knowledge).

\ifthenelse{\boolean{FULL}}{
}{
For the sake of readability, proofs marked with $\ast$ are deferred to \cref{app:proofs}.
}

\subsection{Proof Overview}
\label{subsec:intro:overview}

The construction of alphabet reduction for \prb{Maxmin Binary CSP Reconfiguration} (\cref{thm:ABC-reduct})
is based on that for \prb{Max Binary CSP} due to \citet{dinur2007pcp},
which comprises two partial steps:
The first step is \textbf{robustization}, which replaces
each constraint $\pi_e$ of edge $e$ by
a Boolean circuit $C_e$ that accepts $f \circ g$
if and only if
$f \circ g = \Had(\alpha) \circ \Had(\beta)$ such that
$(\alpha, \beta)$ satisfies $\pi_e$,
where $\Had$ is the Hadamard code (see \cref{sec:pre} for the definition).\footnote{
Though \citet{dinur2007pcp} used
an error-correcting code $\enc \colon \Sigma \to \bbF_2^\ell$ having
\emph{linear dimension} (i.e., $\ell = \bigO(\log |\Sigma|)$),
we can afford to use the Hadamard code for our purpose
because $|\Sigma| = \bigO(1)$.
}
The soundness case ensures that
for ``many'' edges $e$,
the restricted assignment is $\Theta(1)$-far from any satisfying truth assignment to $C_e$.
The second step is \textbf{composition}, which composes
each circuit $C_e$ with
an \emph{assignment tester}~\cite{dinur2007pcp,dinur2006assignment}
(a.k.a.~\emph{PCP of proximity}~\cite{bensasson2006robust}) of constant size
to break down $C_e$ into a system of binary constraints over small alphabet
while sharing the common variables for different circuits.

The main challenge to achieving
alphabet reduction for \prb{Maxmin Binary CSP Reconfiguration} is its robustization.
Simply applying the above robustization procedure to \prb{Maxmin Binary CSP Reconfiguration},
we are required to reconfigure between a pair of codewords, say,
$\Had(\alpha_1)$ and $\Had(\alpha_2)$ for $\alpha_1 \neq \alpha_2$.
Such reconfiguration must
pass through a function $\gtrapprox \frac{1}{4}$-far from the codeword and thus
from any satisfying truth assignment to the above circuit $C_e$,
sacrificing the perfect completeness.
There is a dilemma that distinct codewords should be far from each other,
yet they need to be reconfigurable with each other.
One might thus think of enforcing $C_e$ to accept
functions that are $\frac{1}{4}$-close to the codeword.
Unfortunately, this modification reduces the robustness to $o(1)$ in the soundness case,
as shown in an example below
(see also \cref{eg:robust:detail}).
This explains why robustization for \prb{Maxmin Binary CSP Reconfiguration} is nontrivial.

\begin{example}[Failed attempt]
\label{eg:robust:rough}
Define a binary constraint
$\pi_e \triangleq \{(\alpha_1, \beta_1), (\alpha_2, \beta_2)\} \subset \Sigma \times \Sigma$
and a Hadamard code $\Had \colon \Sigma \to \bbF_2^\ell$.
Construct a (seemingly promising) circuit
$\tilde{C}_e$
such that $\tilde{C}_e(f \circ g) = 1$ if and only if
\begin{enumerate}
    \item both $f$ and $g$ are $\frac{1}{4}$-close to some Hadamard codewords;
    \item if $f$ and $g$ are $\frac{1}{4}$-close to $\Had(\alpha)$ and $\Had(\beta)$, respectively,
    then $(\alpha, \beta)$ must satisfy $\pi_e$.
\end{enumerate}
Then, the following issue arises: even if
$f$ is closest to $\Had(\alpha)$ and
$g$ is closest to $\Had(\beta)$
such that $(\alpha, \beta) \notin \pi$,
we cannot exclude the possibility that
$f \circ g$ is $o(1)$-close to some satisfying truth assignment to $\tilde{C}_e$.
Suppose $f$ is $\frac{1}{4}$-close to both $\Had(\alpha_1)$ and $\Had(\alpha_2)$ and
$g$ is $\frac{1}{4}$-close to both $\Had(\beta_1)$ and $\Had(\beta_2)$.
Changing particular two bits of $f \circ g$, we obtain
$f^\star \circ g^\star$ that is
$\left(\frac{1}{4}-\frac{1}{\ell}\right)$-close to $\Had(\alpha_1) \circ \Had(\beta_1)$.
Since $\tilde{C}(f^\star \circ g^\star) = 1$,
$f \circ g$ is $\frac{1}{\ell}$-close to a satisfying truth assignment to $\tilde{C}_e$.
\lipicsEnd
\end{example}\noindent
The crux of a reconfiguration analogue of robustization is
what we call the \emph{reconfigurability of Hadamard codes}:

\begin{lemma}[Reconfigurability of Hadamard codes; informal; see \cref{lem:Hadmard-reconf}]
There exists a universal constant $\delta_0 \in (0,1)$ such that
for any $n \geq 9$ and $\vec{\alpha} \neq \vec{\beta} \in \bbF_2^n$,
there exists a reconfiguration sequence
from $\Had(\vec{\alpha})$ to $\Had(\vec{\beta})$
such that every function in it is 
\begin{itemize}
    \item $\frac{1}{4}$-close to either 
    $\Had(\vec{\alpha})$ or $\Had(\vec{\beta})$, and
    \item $\left(\frac{1}{4}+\delta_0 \right)$-far from $\Had(\vec{\gamma})$
    for every $\vec{\gamma} \neq \vec{\alpha},\vec{\beta}$.
\end{itemize}
\end{lemma}\noindent
\cref{lem:Hadmard-reconf} enables us to reconfigure between a pair of codewords,
avoiding getting too (say, $\frac{1}{4}+\delta_0$) close to the other codewords.
The existence of such a reconfiguration sequence is shown by a simple application of
the structural property of a triple of distinct Hadamard codewords and
the probabilistic method.
\cref{lem:Hadmard-reconf} is still nontrivial in that
it does not hold if $n=3$ (see \cref{obs:Hadamard-reconf-fail}).
Using the reconfigurability of Hadamard codes,
we implement alphabet reduction of \prb{Maxmin Binary CSP Reconfiguration} as follows:

\begin{itemize}
\item \textbf{Robustization} (\cref{lem:robust}):
Convert a binary constraint $\pi_e$ for edge $e$ into a circuit $C_e$
such that $C_e(f \circ g) = 1$ if and only if
\begin{enumerate}
    \item both $f$ and $g$ are $\frac{1}{4}$-close to some Hadamard codewords;
    \item if $f$ and $g$ are
    \colorbox{yellow!75!white}{$\left(\frac{1}{4}+\frac{\delta_0}{2}\right)$-close}
    to $\Had(\alpha)$ and $\Had(\beta)$, respectively,
    then $(\alpha, \beta)$ must satisfy $\pi_e$.
\end{enumerate}
(The difference from $\tilde{C}_e$ of \cref{eg:robust:rough} is \colorbox{yellow!75!white}{highlighted}.)
Consider $\pi_e$, $f$, and $g$ appearing in \cref{eg:robust:rough} again for the soundness case.
Suppose $C_e$ is constructed from $\pi_e$.
To make $f \circ g$ to satisfy $C_e$,
we must modify them so that $f$ and $g$ are $\left(\frac{1}{4}+\frac{\delta_0}{2}\right)$-far from
$\Had(\alpha_2)$ and $\Had(\beta_2)$
(or $\Had(\alpha_1)$ and $\Had(\beta_1)$), respectively; namely,
$f \circ g$ is $\frac{\delta_0}{2}$-far from any satisfying truth assignment to $C_e$.
Even though $C_e$ demands stricter conditions than $\tilde{C}_e$ of \cref{eg:robust:rough},
the perfect completeness can be derived using \cref{lem:Hadmard-reconf}.

\item \textbf{Composition} (\cref{prp:composition}):
Just feeding each circuit $C_e$ to an assignment tester $\calP$ of \cite{dinur2007pcp}
breaks the perfect completeness; instead,
we apply $\calP$ to $C_e$ \emph{twice} to create
twins of binary constraint systems sharing the input variables to $C_e$.
Our $4$-query verifier then
picks a pair of edges from each of the twins uniformly at random, and
accepts if either of them is satisfied,
which may be thought of as \emph{rectangular PCPs} \cite{bhangale2020rigid}.
This kind of redundancy is crucial for ensuring the perfect completeness of reconfiguration problems.
On the other hand, if $\delta$-fraction of the edges are unsatisfied in both of the twins,
the verifier rejects with probability $\delta^2$ owing to its rectangularity.
\end{itemize}

In the language of probabilistic proofs,
the above alphabet reduction can be thought of as
a composition of
\emph{probabilistically checkable reconfiguration proofs} (PCRPs)
due to \citet{hirahara2024probabilistically},
where an outer PCRP is \prb{Gap Binary CSP Reconfiguration} and
an inner PC(R)P is an assignment tester.
To make the outer PCRP enjoy a reconfiguration analogue of the robustness as in \cref{lem:robust},
we
replace each variable by a block of bits and
modify the original circuit associated with each edge $e$ (i.e., binary constraint $\pi_e$) appropriately
so as to reflect the reconfigurability of Hadamard codes.
% \cite[Lemma~3.6]{dinur2006assignment}\cite[Lemma~2.13]{bensasson2006robust}

\subsection{Towards Dinur's Style Proof of RIH?}
\label{subsec:intro:dinur}
Given
degree reduction \cite{ohsaka2023gap},
gap amplification \cite{ohsaka2024gap}, and
alphabet reduction (this paper) for 
\prb{Maxmin Binary CSP Reconfiguration},
one might think of proving RIH
imitating \citeauthor{dinur2007pcp}'s proof of the PCP theorem~\cite{dinur2007pcp}.
Though RIH has been proven by \citet{karthik2023inapproximability,hirahara2024probabilistically}
(see also \cref{subsec:intro:related}),
such a different proof is still useful in a sense that
it would be more combinatorial and rely only on simple tools.
Unfortunately,
merely putting them together does not work as expected because
some of the reductions are only \emph{gap-preserving},
which requires that
there is already a constant gap $\epsilon \in (0,1)$ between completeness and soundness,
and thus weaker than those of \citet{dinur2007pcp}.
Consider for example degree reduction of \prb{Maxmin Binary CSP Reconfiguration}.
Unlike {Papadimitriou--Yannakakis}'s degree reduction for \prb{Max Binary CSP} \cite{papadimitriou1991optimization},
\citeauthor{ohsaka2023gap}'s degree reduction \cite{ohsaka2023gap} uses
near-Ramanujan graphs \cite{alon2021explicit,mohanty2021explicit} of degree $\Theta(\epsilon^{-2})$.
Since we need to begin gap amplification with $\epsilon = (\text{\# of edges})^{-1} = o(1)$,
applying the degree reduction step of \cite{ohsaka2023gap}
results in a superconstant degree,
failing to reduce the degree of \prb{Maxmin Binary CSP Reconfiguration}.
Gap amplification of \citet{ohsaka2024gap}
also relies on the assumption that the gap value is a constant, see \cite[Claim 3.7]{ohsaka2024gap}.
Note that alphabet reduction in the present study works for any subconstant gap.

\subsection{Additional Related Work}
\label{subsec:intro:related}
In \cite{ito2011complexity},
\NP-hardness of approximation is shown for
optimization versions of \prb{Clique Reconfiguration} and \prb{SAT Reconfiguration}
using \NP-hardness of approximating
\prb{Max Clique}~\cite{hastad1999clique} and \prb{Max SAT}~\cite{hastad2001some}, respectively.
Other reconfiguration problems whose approximability was investigated include
\prb{Subset Sum Reconfiguration}, which admits a PTAS \cite{ito2014approximability} and
\prb{Submodular Reconfiguration}, which admits a constant-factor approximation \cite{ohsaka2022reconfiguration}.
It is known that
a naive parallel repetition for \prb{Maxmin Binary CSP Reconfiguration} fails to decrease the soundness error \cite{ohsaka2023approximate} unlike the parallel repetition theorem due to \citet{raz1998parallel};
in fact, \prb{Maxmin Binary CSP Reconfiguration} is approximable within a factor of nearly $\frac{1}{4}$ \cite{ohsaka2023approximate}
while \NP-hard to approximate within a factor better than $\frac{3}{4}$ \cite{ohsaka2024gap}.
\citet{karthik2023inapproximability} demonstrate matching lower and upper bounds, i.e.,
\NP-hardness of $\left(\frac{1}{2}+\epsilon\right)$-factor approximation and
a $\left(\frac{1}{2}-\epsilon\right)$-factor approximation algorithm
for every $\epsilon \in \left(0,\frac{1}{2}\right)$.

Very recently, \citet{karthik2023inapproximability,hirahara2024probabilistically}
independently announced the proof of RIH;
in particular,
reconfiguration problems listed in \cref{cor:inapprox-reconf} are \emph{unconditionally} \PSPACE-hard to approximate.
Our alphabet reduction is still meaningful,
which makes $\epsilon_0$ and $W_0$ of \cref{cor:ABC-reduct}
independent of the soundness error, query complexity, and alphabet size of the verifier of 
\cite{karthik2023inapproximability,hirahara2024probabilistically}.
Moreover, \cref{thm:ABC-reduct} can be used to 
efficiently reduce the alphabet size of ($2$-query) probabilistically checkable reconfiguration proofs
due to \citet{hirahara2024probabilistically}.

The \emph{overlap gap property} 
\cite{gamarnik2021overlap,achlioptas2011solution,mezard2005clustering,gamarnik2017limits,wein2021optimal}
refers to the separation phenomena of the overlaps between 
near-optimal solutions on random instance, which implies 
approximate reconfigurability; see also \cite{ohsaka2024gap}.

\section{Preliminaries}
\label{sec:pre}

\subsection{Notations}
For a nonnegative integer $n \in \bbN$,
let $ [n] \triangleq \{1, 2, \ldots, n\} $.
Denote by $\mathfrak{S}_n$ the set of all permutations over $[n]$.
A \emph{sequence} $\scrS$ of a finite number of objects $S^{(1)}, \ldots, S^{(\TTT)}$
is denoted by $( S^{(1)}, \ldots, S^{(\TTT)} )$, and
we write $S^{(\ttt)} \in \scrS$ to indicate that $S^{(\ttt)}$ appears in $\scrS$.
The symbol $\circ$ stands for a concatenation of two strings,
$\langle \cdot, \cdot \rangle$ for the inner product,
$\bbF_2 = \{0,1\}$ for the finite field with two elements.
We use $\uplus$ to emphasize that the union is taken over disjoint sets.
Let $\Sigma$ be a finite set called \emph{alphabet}.
For a length-$n$ string $f \in \Sigma^n$ and index set $I \subseteq [n]$,
we use $f|_I$ to denote the restriction of $f$ to $I$.
The \emph{relative distance} between two strings $f,g \in \Sigma^n$,
denoted $\Delta(f,g)$,
is defined as the fraction of positions on which $f$ and $g$ differ; namely,
$
    \Delta(f,g)
    = \Pr_{i \sim [n]}[f_i \neq g_i]
    = \frac{|\{i \in [n] \mid f_i \neq g_i \}|}{n}.
$
We say that
$f$ is \emph{$\epsilon$-close} to $g$ if 
$\Delta(f,g) \leq \epsilon$
and \emph{$\epsilon$-far} from $g$ if
$\Delta(f,g) > \epsilon$.
For a set of strings $S \subseteq \Sigma^n$, analogous notions are used; e.g.,
$\Delta(f, S) \triangleq \min_{g \in S} \Delta(f,g)$ and
$f$ is $\epsilon$-close to $S$ if $\Delta(f,S) \leq \epsilon$.
For a string $\vec{\alpha} \in \bbF_2^n$,
its \emph{Hadamard code} is defined as 
a function $\Had(\vec{\alpha}) \colon \bbF_2^n \to \bbF_2$ such that
$
    \Had(\vec{\alpha})(\vec{x}) = \langle \vec{\alpha}, \vec{x} \rangle
    \text{ for all } \vec{x} \in \bbF_2^n.
$
We call $\Had(\vec{\alpha})$ for each $\vec{\alpha}$ a \emph{codeword} of the Hadamard code, and
write $\Had(\cdot)$ for the set of all codewords.
Note that the relative distance between any pair of distinct codewords of $\Had(\cdot)$ is $\frac{1}{2}$; i.e.,
$\Delta(\Had(\vec{\alpha}), \Had(\vec{\beta})) = \frac{1}{2}$ for all $\vec{\alpha} \neq \vec{\beta} \in \bbF_2^n$.

\subsection{Constraint Satisfaction Problem and Reconfigurability}
We introduce reconfiguration problems on constraint satisfaction.
The notion of constraint graphs is first introduced.

\begin{definition}
A \emph{$q$-ary constraint graph} is defined as a tuple $G=(V,E,\Sigma,\Pi)$ such that
$(V,E)$ is a $q$-uniform hypergraph called the \emph{underlying graph},
$\Sigma$ is a finite set called the \emph{alphabet}, and
$\Pi = (\pi_e)_{e \in E}$ is a collection of $q$-ary \emph{constraints}, where
each constraint $\pi_e \subseteq \Sigma^e$ is a set of $q$-tuples of acceptable values that $q$ vertices in $e$ can take.
\lipicsEnd
\end{definition}
For an \emph{assignment} $\psi \colon V \to \Sigma$,
we say that $\psi$ \emph{satisfies} hyperedge $e = \{v_1, \ldots, v_q\} \in E$ (or constraint $\pi_e$) if
$\psi(e) \triangleq (\psi(v_1), \ldots, \psi(v_q)) \in \pi_e$, and
$\psi$ \emph{satisfies} $G$ if it satisfies all hyperedges of $G$.
For two satisfying assignments $\psi^\ini$ and $\psi^\tar$ for $G$,
a \emph{reconfiguration sequence from $\psi^\ini$ to $\psi^\tar$} over $\Sigma^V$
is any sequence
$( \psi^{(1)}, \ldots \psi^{(\TTT)} )$ such that
$\psi^{(1)} = \psi^\ini$,
$\psi^{(\TTT)} = \psi^\tar$, and
every two neighboring assignments $\psi^{(\ttt)}$ and $\psi^{(\ttt+1)}$ differ in at most one vertex.
In the \prb{$q$-CSP Reconfiguration} problem,
for a $q$-ary constraint graph $G$ and its two satisfying assignments $\psi^\ini$ and $\psi^\tar$,
we are asked to decide if there is a reconfiguration sequence of satisfying assignments for $G$ from $\psi^\ini$ to $\psi^\tar$.
Hereafter, \prb{BCSP} stands for \prb{$2$-CSP}, and
the suffix ``$_W$'' designates the restricted case that the alphabet size $|\Sigma|$ is integer $W \in \bbN$.

Subsequently,
we formulate an optimization version of \prb{$q$-CSP Reconfiguration} \cite{ito2011complexity,ohsaka2023gap},
which allows going through non-satisfying assignments.
For a constraint graph $G=(V,E,\Sigma,\Pi)$ and an assignment $\psi \colon V \to \Sigma$,
its \emph{value} is defined as the fraction of edges of $G$ satisfied by $\psi$; namely,
\begin{align}
    \val_G(\psi) \triangleq \frac{1}{|E|} \cdot
    \left|\Bigl\{e \in E \Bigm| \psi \text{ satisfies } e \Bigr\}\right|.
\end{align}
For a reconfiguration sequence
$\seqpsi = ( \psi^{(1)}, \ldots, \psi^{(\TTT)} )$ of assignments,
let $\val_G(\seqpsi)$ denote the \emph{minimum fraction} of satisfied edges
over all $\psi^{(\ttt)}$'s in $\seqpsi$; namely,
\begin{align}
    \val_G(\seqpsi) \triangleq \min_{\psi^{(\ttt)} \in \seqpsi} \val_G(\psi^{(\ttt)}).
\end{align}
In \prb{Maxmin $q$-CSP Reconfiguration},
we wish to maximize $\val_G(\seqpsi)$ subject to $\seqpsi = ( \psi^\ini, \ldots, \psi^\tar )$.
For two assignments $\psi^\ini, \psi^\tar \colon V \to \Sigma$ for $G$,
let $\val_G(\psi^\ini \reco \psi^\tar)$ denote the maximum value of $\val_G(\seqpsi)$
over all possible reconfiguration sequences $\seqpsi$ from $\psi^\ini$ to $\psi^\tar$; namely,
\begin{align}
    \val_G(\psi^\ini \reco \psi^\tar)
    \triangleq \max_{\seqpsi = ( \psi^\ini, \ldots, \psi^\tar )} \val_G(\seqpsi)
    = \max_{\seqpsi = ( \psi^\ini, \ldots, \psi^\tar )} \min_{\psi^{(\ttt)} \in \seqpsi} \val_G(\psi^{(\ttt)}).
\end{align}
The gap version of \prb{Maxmin $q$-CSP Reconfiguration} is defined as follows.

\begin{problem}
\label{prb:gap-CSP}
For every numbers $0 \leq s \leq c \leq 1$ and integer $q \in \bbN$,
\prb{Gap$_{c,s}$ $q$-CSP Reconfiguration} requests to determine
for a $q$-ary constraint graph $G$ and its two assignments $\psi^\ini$ and $\psi^\tar$,
whether
$\val_G(\psi^\ini \reco \psi^\tar) \geq c$ (the input is a \YES instance) or
$\val_G(\psi^\ini \reco \psi^\tar) < s$ (the input is a \NO instance).
Here, $c$ and $s$ are respectively called \emph{completeness} and \emph{soundness}.
\lipicsEnd
\end{problem}\noindent
We can assume $\psi^\ini$ and $\psi^\tar$ satisfy $G$ whenever $c=1$.
The \emph{Reconfiguration Inapproximability Hypothesis} (RIH)~\cite{ohsaka2023gap}
postulates that 
\prb{Gap$_{1, 1-\epsilon}$ $q$-CSP$_W$ Reconfiguration} is \PSPACE-hard
for some $\epsilon \in (0,1)$ and $q, W \in \bbN$,
which has been recently proven by \citet{karthik2023inapproximability,hirahara2024probabilistically}.

\section{Alphabet Reduction for Maxmin BCSP Reconfiguration}
\label{sec:main}

In this section, we prove the main result of this paper, i.e.,
an explicit construction of \emph{alphabet reduction} for \prb{Maxmin BCSP Reconfiguration},
as formally stated below.

\begin{theorem}[Alphabet reduction]
\label{thm:ABC-reduct}
There exist
universal constants $W_0 \in \bbN$ and $\kappa \in (0,1)$, and
a polynomial-time algorithm $\calA$ that takes an instance
$(G,\psi^\ini,\psi^\tar)$
of \prb{Maxmin BCSP$_W$ Reconfiguration}
with alphabet size $W \in \bbN$
and
produces an instance
$(G',\psi'^\ini,\psi'^\tar)$
of \prb{Maxmin BCSP$_{W_0}$ Reconfiguration}
with alphabet size $W_0$
such that the following hold\textup{:}
\begin{itemize}
    \item \textup{(}Perfect completeness\textup{)}
    If $\val_G(\psi^\ini \reco \psi^\tar) = 1$, then
    $\val_{G'}(\psi'^\ini \reco \psi'^\tar) = 1$.
    \item \textup{(}Soundness\textup{)} 
    If $\val_G(\psi^\ini \reco \psi^\tar) < 1-\epsilon$, then
    $\val_{G'}(\psi'^\ini \reco \psi'^\tar) < 1- \kappa \cdot \epsilon$.
\end{itemize}
In particular,
for every $\epsilon \in (0,1)$ and $W \in \bbN$,
$\calA$ is a gap-preserving reduction from
\prb{Gap$_{1,1-\epsilon}$ BCSP$_W$ Reconfiguration} to
\prb{Gap$_{1,1-\kappa \cdot \epsilon}$ BCSP$_{W_0}$ Reconfiguration}.
\end{theorem}

The remainder of this section is organized as follows:
\cref{subsec:main:Hadamard} introduces and proves the reconfigurability of Hadamard codes,
which will be applied to robustization of \prb{Maxmin BCSP Reconfiguration} in \cref{subsec:main:robust}.
Subsequently,
\cref{subsec:main:composition} composes the assignment tester of \cite{dinur2007pcp,odonnell2014analysis}
into \prb{Circuit SAT Reconfiguration}, concluding the proof of \cref{thm:ABC-reduct}.

\subsection{Reconfigurability of Hadamard Codes}
\label{subsec:main:Hadamard}

Here, we prove the reconfigurability of Hadamard codewords.
A \emph{reconfiguration sequence from $f^\ini$ to $f^\tar$} over $\bbF_2^N$
is a sequence $( f^{(1)}, \ldots, f^{(\TTT)} )$ such that
$f^{(1)} = f^\ini$,
$f^{(\TTT)} = f^\tar$, and
every two neighboring functions $f^{(\ttt)}$ and $f^{(\ttt+1)}$ differ in at most one bit.

\begin{lemma}[Reconfigurability of Hadamard codes]
\label{lem:Hadmard-reconf}
Let $n$ be a positive integer at least $9$,
    $\delta_0 \triangleq \frac{1}{400}$ be a universal constant, and
    $\vec{\alpha}, \vec{\beta} \in \bbF_2^n$ be two distinct strings.
    Then, there exists a reconfiguration sequence 
    $\Pi = ( \Had(\vec{\alpha}), \ldots, \Had(\vec{\beta}) )$
    from $\Had(\vec{\alpha})$
    to $\Had(\vec{\beta})$
    such that
    for every string $\vec{\gamma} \in \bbF_2^n \setminus \{\vec{\alpha}, \vec{\beta}\}$ and
    every function $f \colon \bbF_2^n \to \bbF$ in $\Pi$,
    \begin{align}
        \min\Bigl\{ \Delta(f, \Had(\vec{\alpha})), \Delta(f, \Had(\vec{\beta})) \Bigr\} & \leq \frac{1}{4}, \\
        \Delta(f, \Had(\vec{\gamma})) & > \frac{1}{4} + \delta_0.
    \end{align}
\end{lemma}

Before going to its proof,
we remark that \cref{lem:Hadmard-reconf} \emph{does not} hold if $n=3$.
\ifthenelse{\boolean{FULL}}{
\begin{observation}
}{
\begin{observation}[$\ast$]
}
\label{obs:Hadamard-reconf-fail}
For $n=3$ and $\vec{\alpha} \neq \vec{\beta} \in \bbF_2^n$, 
let $\Pi$ be a reconfiguration sequence
from $\Had(\vec{\alpha})$ to $\Had(\vec{\beta})$
such that 
$\min\{\Delta(f, \Had(\vec{\alpha})), \Delta(f, \Had(\vec{\beta})) \} \leq \frac{1}{4}$
for every function $f$ in $\Pi$.
Then, $\Pi$ contains a function $f^\circ \colon \bbF_2^n \to \bbF_2$ such that
$\Delta(f^\circ, \Had(\vec{\gamma})) \leq \frac{1}{4}$
for some $\vec{\gamma} \in \bbF_2^n \setminus \{\vec{\alpha}, \vec{\beta}\}$.
\end{observation}
\ifthenelse{\boolean{FULL}}{
\begin{table}
    \newcommand{\D}{\cellcolor{blue!30}}
    \newcommand{\pick}{\cellcolor{red!30}}
    \centering
    \begin{tabular}{c|cccccccc}
        \toprule
        position $\vec{x}$ & $000$ & \D$001$ & $010$ & $100$ & $110$ & \D$101$ & \D$011$ & \D$111$ \\
        \midrule
        $\Had(000)(\vec{x})$ & $0$ & \D$0$ & $0$ & $0$ & $0$ & \D$0$ & \D$0$ & \D$0$ \\
        $\Had(001)(\vec{x})$ & $0$ & \D$1$ & $0$ & $0$ & $0$ & \D$1$ & \D$1$ & \D$1$ \\
        \midrule
        $\Had(010)(\vec{x})$ & $0$ & $0$ & $1$ & $0$ & $1$ & $0$ & \pick$1$ & \pick$1$ \\
        $\Had(100)(\vec{x})$ & $0$ & $0$ & $0$ & $1$ & $1$ & \pick$1$ & $0$ & \pick$1$ \\
        $\Had(110)(\vec{x})$ & $0$ & $0$ & $1$ & $1$ & $0$ & \pick$1$ & \pick$1$ & $0$ \\
        $\Had(101)(\vec{x})$ & $0$ & \pick$1$ & $0$ & $1$ & $1$ & $0$ & \pick$1$ & $0$ \\
        $\Had(011)(\vec{x})$ & $0$ & \pick$1$ & $1$ & $0$ & $1$ & \pick$1$ & $0$ & $0$ \\
        $\Had(111)(\vec{x})$ & $0$ & \pick$1$ & $1$ & $1$ & $0$ & $0$ & $0$ & \pick$1$ \\
        \toprule
    \end{tabular}
    \caption{
        Hadamard codewords in the case of $n=3$.
        If a function $f^\circ \colon \bbF_2^n \to \bbF_2$ is obtained from $\Had(000)$ by
        flipping $2$ bits of
        $D = \{\vec{x} \in \bbF_2^n \mid \langle \vec{x}, \vec{\alpha} \rangle \neq \langle \vec{x}, \vec{\beta} \rangle\} = \{001,101,011,111\}$,
        where $\vec{\alpha} = 000$ and $\vec{\beta} = 001$,
        then $f^\circ$ is $\frac{1}{4}$-close to $\Had(\vec{\gamma})$ for some $\vec{\gamma} \neq \vec{\alpha}, \vec{\beta}$.
    }
    \label{tab:Hadamard-reconf-fail}
\end{table}

\ifthenelse{\boolean{FULL}}{
\begin{proof}
}{
\begin{proof}[Proof of \cref{obs:Hadamard-reconf-fail}]
}
Consider that $\vec{\alpha} = 000$ and $\vec{\beta} = 001$;
the other cases can be shown analogously.
Suppose a reconfiguration sequence $\Pi$ from $\Had(\vec{\alpha})$ to $\Had(\vec{\beta})$
satisfies that 
\begin{align}
    \min\Bigl\{\Delta(f, \Had(\vec{\alpha})), \Delta(f, \Had(\vec{\beta})) \Bigr\} \leq \frac{1}{4} \text{ for all } f \in \Pi.
\end{align}
Then, $\Pi$ must contain a function $f^\circ \colon \bbF_2^n \to \bbF_2$ such that
$\Delta(f^\circ, \Had(\vec{\alpha})) = \Delta(f^\circ, \Had(\vec{\beta})) = \frac{1}{4}$.
Such $f^\circ$ is thus obtained from $\Had(\vec{\alpha})$ (or $\Had(\vec{\beta})$)
by flipping $2$ bits of the $4$ bits $D$
on which $\Had(\vec{\alpha})$ and $\Had(\vec{\beta})$ disagree with each other.
Specifically, we have
\begin{align}
    D
    \triangleq \Bigl\{\vec{x} \in \bbF_2^n \Bigm| \langle \vec{x}, \vec{\alpha} \rangle \neq \langle \vec{x}, \vec{\beta} \rangle \Bigr\}
    = \Bigl\{001, 101, 011, 111\Bigr\}.
\end{align}
For any $\vec{x} \in \bbF_2$,
let $\vec{1}_{\vec{x}} \colon \bbF_2^n \to \bbF_2$ denote a function such that
$\vec{1}_{\vec{x}}(\vec{y}) = \llbracket \vec{x} = \vec{y} \rrbracket$.
Then, $f^\circ = \Had(\vec{\alpha}) + \vec{1}_{\vec{x}} + \vec{1}_{\vec{y}}$
for some $\vec{x} \neq \vec{y} \in D$.
Observe easily from \cref{tab:Hadamard-reconf-fail} that
\begin{align}
    \forall \vec{x} \neq \vec{y} \in D, \;\;
    \exists \vec{\gamma} \in \bbF_2^n \setminus \{\vec{\alpha}, \vec{\beta}\}, \;\;
    \Delta(\Had(\vec{\alpha}) + \vec{1}_{\vec{x}} + \vec{1}_{\vec{y}}, \Had(\vec{\gamma})) \leq \frac{1}{4},
\end{align}
as desired.
For example, if $\vec{x}=011$ and $\vec{y}=111$, then
$\Delta(\Had(\vec{\alpha}) + \vec{1}_{\vec{x}} + \vec{1}_{\vec{y}}, \Had(010)) \leq \frac{1}{4}$.
\end{proof}

}{
}

To prove \cref{lem:Hadmard-reconf},
we first analyze the partial sum of a random sequence consisting of
an equal number of plus ones and minus ones.

\ifthenelse{\boolean{FULL}}{
\begin{lemma}
}{
\begin{lemma}[$\ast$]
}
\label{lem:random-perm}
Let $N > N_0 \triangleq 100$ be any positive integer,
$\eta_0 \triangleq \frac{1}{100}$, and
$\vec{a} = ( a_1, \ldots, a_{2N} )$ be
a random sequence made up of $N$ plus ones and $N$ minus ones
obtained by applying a random permutation of $\mathfrak{S}_{2N}$ to
$( \underbrace{+1, \ldots, +1}_{N \text{ times}}, \underbrace{-1, \ldots, -1}_{N \text{ times}} )$.
Then, the minimum $k$-partial sum over all $k \in [2N]$\textup{;} i.e.,
\begin{align}
    \argmin_{1 \leq k \leq 2N} \sum_{1 \leq i \leq k} a_i
    = \argmin_{1 \leq k \leq 2N} \sum_{k+1 \leq i \leq 2N} a_i,
\end{align}
is at most $- (1-\eta_0) N = -0.99N$ with probability at most 
$0.9^N$.
\end{lemma}
\ifthenelse{\boolean{FULL}}{
\ifthenelse{\boolean{FULL}}{
\begin{proof}
}{
\begin{proof}[Proof of \cref{lem:random-perm}]
}
Observe first that the $k$-partial sum may be at most $-(1-\eta_0)N$ only if
\begin{align}
    (1-\eta_0) N \leq k \leq (1+\eta_0) N.
\end{align}
For such fixed $k$,
we will bound the number of permutations in $\mathfrak{S}_{2N}$
inducing that
$\sum_{1 \leq i \leq k} a_i \leq -(1-\eta_0)N$.
Let $N_+$ and $N_-$ denote the number of plus ones and minus ones in
$( a_1, \ldots, a_k )$, respectively.
Since $N_+ + N_- = k$, for the $k$-partial sum to be at most $-(1-\eta_0)N$,
it must hold that
\begin{align}
\begin{aligned}
    & \sum_{1 \leq i \leq k} a_i = N_+ - N_- \leq -(1-\eta_0) N \\
    & \implies N_- \geq \frac{1-\eta_0}{2}N + \frac{k}{2}.
\end{aligned}
\end{align}
Note also that $N_- \leq \min\{N, k\}$.
The number of permutations making
$(a_1, \ldots, a_k)$ to include $i$ minus ones for $0 \leq i \leq k$
is equal to
\begin{align}
    {N \choose i} \cdot {N \choose k-i} \cdot k! \cdot (2N-k!),
\end{align}
implying that
the total number of permutations ensuring the desired event
is
\begin{align}
    \sum_{\frac{1-\eta_0}{2}N + \frac{k}{2} \leq i \leq \min\{N,k\}}
    {N \choose i} \cdot {N \choose k-i} \cdot k! \cdot (2N-k!).
\end{align}
Since
$i \geq \frac{1-\eta_0}{2}N + \frac{k}{2} \geq (1-\epsilon)N$ and
$k \leq (1+\eta_0)N$, we bound
\begin{align}
\begin{aligned}
    {N \choose i} \cdot {N \choose k-i}
    & \leq {N \choose (1-\eta_0) N} \cdot {N \choose 2\eta_0 N} \\
    & \leq
        \left(\frac{\rme \cdot N}{(1-\eta_0) N}\right)^{(1-\eta_0)N} \cdot
        \left(\frac{\rme \cdot N}{2\eta_0 N}\right)^{2\eta_0 N} \\
    & = \left(
        \left(\frac{\rme}{1-\eta_0}\right)^{1-\eta_0} \cdot
        \left(\frac{\rme}{2\eta_0}\right)^{2\eta_0}
    \right)^N \\
    & = \left(
        \left(\frac{\rme}{0.99}\right)^{0.99} \cdot
        \left(\frac{\rme}{0.02}\right)^{0.02}
    \right)^N 
    < 3^N,
\end{aligned}
\end{align}
where the second inequality is due to the upper bound that
${n \choose m} < \left(\frac{\rme \cdot n}{m}\right)^m$ for any $n$ and $m$.
Noting that the number of permutations in $\mathfrak{S}_{2N}$ is $(2N)!$,
we obtain
\begin{align}
\begin{aligned}
    & \Pr_{\vec{a}}\left[\sum_{1 \leq i \leq k} a_i < -(1-\eta_0) N\right] \\
    & = \frac{1}{(2N)!}
    \sum_{\frac{1-\eta_0}{2}N + \frac{k}{2} \leq i \leq \min\{k,N\}}
        {N \choose i} \cdot {N \choose k-i} \cdot k! \cdot (2N-k!) \\
    & < \frac{N \cdot 3^N \cdot k! \cdot (2N-k)!}{(2N)!} 
    = \frac{N \cdot 3^N}{{2N \choose k}}.
\end{aligned}
\end{align}

We now bound ${2N \choose k}^{-1}$.
Using the inequality that 
\begin{align}
    \frac{1}{n+1} 2^{n \mathrm{H}\left(\frac{m}{n}\right)} \leq {n \choose m},
\end{align}
where $\mathrm{H}$ is the binary entropy function defined as
\begin{align}
    \mathrm{H}(p) \triangleq -p \log (p) - (1-p) \log (1-p),
\end{align}
we derive
\begin{align}
\begin{aligned}
    {2N \choose k}
    & \geq \min_{(1-\eta_0)N \leq k' \leq (1+\eta_0)N}
    {2N \choose k'}
    \geq {2N \choose 0.99 N} \\
    & \geq \frac{1}{2N+1} 2^{2N \cdot \mathrm{H}\left(\frac{0.99N}{2N}\right)}
    \geq \frac{1}{3N} 4^{0.99N}.
\end{aligned}
\end{align}

Taking a union bound over all possible $k$, we obtain
\begin{align}
\begin{aligned}
    & \Pr_{\vec{a}}\left[\exists k \text{ s.t. } \sum_{1 \leq i \leq k} a_i < -(1-\eta_0)N\right] \\
    & \leq \sum_{(1-\eta_0) N \leq k \leq (1+\eta_0) N}
        \Pr_{\vec{a}}\left[\sum_{1 \leq i \leq k} a_i < -(1-\eta_0)N\right] \\
    & \leq \sum_{(1-\eta_0) N \leq k \leq (1+\eta_0) N}
    3N^2 \cdot 3^N \cdot 4^{-0.99N} \\
    & \leq 6N^3 \cdot 3^N \cdot 4^{-0.99N}
    < 0.9^N & (\text{for all } N > 100).
\end{aligned}
\end{align}
This accomplishes the proof.
\end{proof}

}{
}

Besides,
given the Hadamard codewords of any three distinct strings,
we partition their bits into four equal-sized groups.

\ifthenelse{\boolean{FULL}}{
\begin{claim}
}{
\begin{claim}[$\ast$]
}
\label{clm:alpha-beta-gamma}
For three distinct vectors $\vec{\alpha}, \vec{\beta}, \vec{\gamma} \in \bbF_2^n$, the following hold:
\begin{align}
\begin{aligned}
    \Pr_{\vec{x} \in \bbF_2^n}\Bigl[
        \langle \vec{\alpha}, \vec{x} \rangle \neq \langle \vec{\beta}, \vec{x} \rangle = \langle \vec{\gamma}, \vec{x} \rangle
    \Bigr]
    = \frac{1}{4}, \quad
    \Pr_{\vec{x} \in \bbF_2^n}\Bigl[
        \langle \vec{\beta}, \vec{x} \rangle \neq \langle \vec{\gamma}, \vec{x} \rangle = \langle \vec{\alpha}, \vec{x} \rangle
    \Bigr]
    = \frac{1}{4}, \\
    \Pr_{\vec{x} \in \bbF_2^n}\Bigl[
        \langle \vec{\gamma}, \vec{x} \rangle \neq \langle \vec{\alpha}, \vec{x} \rangle = \langle \vec{\beta}, \vec{x} \rangle
    \Bigr]
    = \frac{1}{4}, \quad
    \Pr_{\vec{x} \in \bbF_2^n}\Bigl[
        \langle \vec{\alpha}, \vec{x} \rangle = \langle \vec{\beta}, \vec{x} \rangle = \langle \vec{\gamma}, \vec{x} \rangle
    \Bigr]
    = \frac{1}{4}.
\end{aligned}
\end{align}
\end{claim}
\ifthenelse{\boolean{FULL}}{
\ifthenelse{\boolean{FULL}}{
\begin{proof}
}{
\begin{proof}[Proof of \cref{clm:alpha-beta-gamma}]
}
Observe first that for any distinct vectors
$\vec{v}_1 \neq \vec{v}_2 \in \bbF_2^n$ and
any $b_1, b_2 \in \bbF_2$, 
\begin{align}
\label{eq:alpha-beta-gamma:quarter}
    \Pr_{\vec{x} \in \bbF_2^n}\Bigl[
        \langle \vec{v}_1, \vec{x} \rangle = b_1 \wedge
        \langle \vec{v}_2, \vec{x} \rangle = b_2
    \Bigr]
    = \frac{1}{4}.
\end{align}
Indeed, it holds that
\begin{align}
\begin{aligned}
    & \Pr_{\vec{x} \in \bbF_2^n}\Bigl[
        \langle \vec{v}_1, \vec{x} \rangle = b_1 \wedge
        \langle \vec{v}_2, \vec{x} \rangle = b_2
    \Bigr]
    +
    \Pr_{\vec{x} \in \bbF_2^n}\Bigl[
        \langle \vec{v}_1, \vec{x} \rangle = b_1 \wedge
        \langle \vec{v}_2, \vec{x} \rangle = 1-b_2
    \Bigr]
    = \frac{1}{2} \\
    & \Pr_{\vec{x} \in \bbF_2^n}\Bigl[
        \langle \vec{v}_1, \vec{x} \rangle = 1-b_1 \wedge
        \langle \vec{v}_2, \vec{x} \rangle = 1-b_2
    \Bigr]
    +
    \Pr_{\vec{x} \in \bbF_2^n}\Bigl[
        \langle \vec{v}_1, \vec{x} \rangle = b_1 \wedge
        \langle \vec{v}_2, \vec{x} \rangle = 1-b_2
    \Bigr]
    = \frac{1}{2} \\
    & \implies
    \Pr_{\vec{x} \in \bbF_2^n}\Bigl[
        \langle \vec{v}_1, \vec{x} \rangle = b_1 \wedge
        \langle \vec{v}_2, \vec{x} \rangle = b_2
    \Bigr]
    =
    \Pr_{\vec{x} \in \bbF_2^n}\Bigl[
        \langle \vec{v}_1, \vec{x} \rangle = 1-b_1 \wedge
        \langle \vec{v}_2, \vec{x} \rangle = 1-b_2
    \Bigr],
\end{aligned}
\end{align}
and
\begin{align}
\begin{aligned}
    & \Pr_{\vec{x} \in \bbF_2^n}\Bigl[
        \langle \vec{v}_1, \vec{x} \rangle = b_1 \wedge
        \langle \vec{v}_2, \vec{x} \rangle = b_2
    \Bigr]
    +
    \Pr_{\vec{x} \in \bbF_2^n}\Bigl[
        \langle \vec{v}_1, \vec{x} \rangle = 1-b_1 \wedge
        \langle \vec{v}_2, \vec{x} \rangle = 1-b_2
    \Bigr] \\
    & = \Pr_{\vec{x} \in \bbF_2^n}\Bigl[
        \langle \vec{v}_1 + \vec{v}_2, \vec{x} \rangle
        = b_1+b_2
    \Bigr]
    = \frac{1}{2},
\end{aligned}
\end{align}
which implies \cref{eq:alpha-beta-gamma:quarter}.
Since $\vec{\alpha} + \vec{\beta} \neq \vec{\beta} + \vec{\gamma}$,
we use \cref{eq:alpha-beta-gamma:quarter} to obtain
\begin{align}
\begin{aligned}
    & \Pr_{\vec{x} \in \bbF_2^n}\Bigl[
        \langle \vec{\alpha}, \vec{x} \rangle =
        \langle \vec{\beta}, \vec{x} \rangle =
        \langle \vec{\gamma}, \vec{x} \rangle
    \Bigr]
    =     \Pr_{\vec{x} \in \bbF_2^n}\Bigl[
        \langle \vec{\alpha} + \vec{\beta}, \vec{x} \rangle = 0 \wedge
        \langle \vec{\beta} + \vec{\gamma}, \vec{x} \rangle = 0
    \Bigr]
    = \frac{1}{4}, \\
    & \Pr_{\vec{x} \in \bbF_2^n}\Bigl[
        \langle \vec{\alpha}, \vec{x} \rangle \neq
        \langle \vec{\beta}, \vec{x} \rangle =
        \langle \vec{\gamma}, \vec{x} \rangle
    \Bigr]
    =     \Pr_{\vec{x} \in \bbF_2^n}\Bigl[
        \langle \vec{\alpha} + \vec{\beta}, \vec{x} \rangle = 1 \wedge
        \langle \vec{\beta} + \vec{\gamma}, \vec{x} \rangle = 0
    \Bigr]
    = \frac{1}{4}.
\end{aligned}
\end{align}
The remaining two cases can be shown similarly, as desired.
\end{proof}

}{
}

Using \cref{lem:random-perm,clm:alpha-beta-gamma},
we now prove \cref{lem:Hadmard-reconf}.

\begin{proof}[Proof of \cref{lem:Hadmard-reconf}]
Fix two strings $\vec{\alpha} \neq \vec{\beta} \in \bbF_2^n$ for $n \geq 9$.
Let $D \subset \bbF_2^n$ be a set of strings on which 
$\Had(\vec{\alpha})$ and $\Had(\vec{\beta})$ disagree with each other; namely,
\begin{align}
    D \triangleq \Bigl\{
        \vec{x} \in \bbF_2^n
        \Bigm|
        \langle \vec{\alpha}, \vec{x} \rangle \neq \langle \vec{\beta}, \vec{x} \rangle
    \Bigr\}.
\end{align}
The random subsum principle ensures $|D| = 2^{n-1}$ (cf.~\cite[Claim~A.31]{arora2009computational}). 
Consider a random reconfiguration sequence
$\Pi = ( \Had(\vec{\alpha}), \ldots, \Had(\vec{\beta}) )$
obtained by the following procedure:
\begin{itembox}[l]{\textbf{Random reconfiguration $\Pi$ from $\Had(\vec{\alpha})$ to $\Had(\vec{\beta})$.}}
\begin{algorithmic}[1]
    \State $(\vec{x}_1, \ldots, \vec{x}_{2^{n-1}}) \leftarrow$
        a sequence obtained by applying 
        a random permutation of $\mathfrak{S}_{2^{n-1}}$ to $D$.
    \For{$i = 1 \textbf{ to } 2^{n-1}$}
        \State flip \nth{$\vec{x}_i$} entry of the current function.
    \EndFor
\end{algorithmic}
\end{itembox}
Observe easily that
any intermediate function of $\Pi$ is always $\frac{1}{4}$-close to
either $\Had(\vec{\alpha})$ or $\Had(\vec{\beta})$.
Fix any string $\vec{\gamma} \in \bbF_2^n \setminus \{\vec{\alpha}, \vec{\beta}\}$.
We would like to show that
with probability at least $1 - 0.9^{2^{n-2}}$,
every function of $\Pi$ is $\left(\frac{1}{4} + \delta_0\right)$-far from $\Had(\vec{\gamma})$; i.e.,
\begin{align}
    \Delta(\Had(\vec{\gamma}), \Pi)
    \triangleq \min_{f \in \Pi} \Delta(\Had(\vec{\gamma}), f)
    > \frac{1}{4} + \delta_0.
\end{align}
By \cref{clm:alpha-beta-gamma},
there exists a partition $(P_\alpha, P_\beta, P_\gamma, P_=)$ of $\bbF_2^n$ such that
$|P_\alpha| = |P_\beta| = |P_\gamma| = |P_=| = 2^{n-2}$ and
\begin{align}
\begin{aligned}
    \langle\vec{\alpha}, \vec{x}\rangle \neq \langle\vec{\beta}, \vec{x}\rangle = \langle\vec{\gamma}, \vec{x}\rangle
    \text{  for all  } \vec{x} \in P_\alpha, \quad
    \langle\vec{\beta}, \vec{x}\rangle \neq \langle\vec{\gamma}, \vec{x}\rangle = \langle\vec{\alpha}, \vec{x}\rangle
    \text{  for all  } \vec{x} \in P_\beta, \\
    \langle\vec{\gamma}, \vec{x}\rangle \neq \langle\vec{\alpha}, \vec{x}\rangle = \langle\vec{\beta}, \vec{x}\rangle
    \text{  for all  } \vec{x} \in P_\gamma, \quad
    \langle\vec{\alpha}, \vec{x}\rangle =    \langle\vec{\beta}, \vec{x}\rangle = \langle\vec{\gamma}, \vec{x}\rangle
    \text{  for all  } \vec{x} \in P_=.
\end{aligned}
\end{align}
See also \cref{tab:alpha-beta-gamma}.
Here, we always have $P_\alpha \uplus P_\beta = D$
(though $P_\alpha$ and $P_\beta$ themselves depend on $\vec{\gamma}$).

For any intermediate function $f \colon \bbF_2^n \to \bbF_2$ of $\Pi$,
if its entry on $P_\alpha$ is flipped,
its Hamming distance to $\Had(\vec{\gamma})$ must decrease by $1$, whereas
if its entry on $P_\beta$ is flipped,
its Hamming distance to $\Had(\vec{\gamma})$ must increase by $1$;
see also \cref{fig:Hadmard-reconf}.
Since $|P_\alpha| = |P_\beta| = 2^{n-2}  > 100$ and
$\| \Had(\vec{\alpha}) - \Had(\vec{\gamma})\| = \|\Had(\vec{\beta}) - \Had(\vec{\gamma})\| = 2^{n-1}$,
we can apply \cref{lem:random-perm} with $N = 2^{n-2}$ to conclude that
\begin{align*}
\begin{aligned}
    & \Pr_{\Pi}\left[
        \min_{f \in \Pi} \| \Had(\vec{\gamma}) - \Pi\| \leq 2^{n-1} - 0.99 N
    \right] \leq 0.9^N 
    \\
    \implies
    & \Pr_{\Pi}\left[
        \Delta(\Had(\vec{\gamma}), \Pi) \leq \frac{1}{4} + \frac{1}{400}
    \right] \leq 0.9^{2^{n-2}}.
\end{aligned}
\end{align*}
Taking a union bound over all possible strings $\vec{\gamma} \in \bbF_2^n \setminus \{\vec{\alpha}, \vec{\beta}\}$,
we derive
\begin{align}
\begin{aligned}
    & \Pr_{\Pi}\left[
        \exists \vec{\gamma} \notin \{\vec{\alpha}, \vec{\beta}\}
        \text{ s.t. }
        \Delta(\Had(\vec{\gamma}), \Pi) \leq \frac{1}{4} + \frac{1}{400}
    \right]
    \\
    & \leq \sum_{\vec{\gamma} \notin \{\vec{\alpha}, \vec{\beta}\}}
    \Pr_{\Pi}\left[
        \Delta(\Had(\vec{\gamma}), \Pi) \leq \frac{1}{4} + \frac{1}{400}
    \right]
    < 2^n \cdot 0.9^{2^{n-2}} 
    < 1 & (\text{for all } n \geq 9).
\end{aligned}
\end{align}
Consequently, the probabilistic method guarantees
the existence of
a reconfiguration sequence $\Pi = ( \Had(\vec{\alpha}), \ldots, \Had(\vec{\beta}) )$
that is entirely $\left(\frac{1}{4}+\delta_0\right)$-far from $\Had(\vec{\gamma})$ for every
$\vec{\gamma} \notin \{\vec{\alpha}, \vec{\beta}\}$.
\end{proof}

\begin{figure}
\null\hfill%
\begin{minipage}{0.5\hsize}%
    \centering
    \footnotesize
    \begin{tabular}{ccccccccc}
         \cline{2-9}
         \multicolumn{1}{c|}{$\Had(\vec{\alpha})$} & \multicolumn{4}{c|}{0} & \multicolumn{4}{c|}{1} \\
         \cline{2-9}
         \\
         \cline{2-9}
         \multicolumn{1}{c|}{$\Had(\vec{\beta})$} &
         \multicolumn{2}{c|}{0} & \multicolumn{2}{c|}{1} & \multicolumn{2}{c|}{0} & \multicolumn{2}{c|}{1} \\
         \cline{2-9}
         \\
         \cline{2-9}
         \multicolumn{1}{c|}{$\Had(\vec{\gamma})$} &
         \multicolumn{1}{c|}{0} & \multicolumn{1}{c|}{1} & \multicolumn{1}{c|}{0} & \multicolumn{1}{c|}{1} & \multicolumn{1}{c|}{0} & \multicolumn{1}{c|}{1} & \multicolumn{1}{c|}{0} & \multicolumn{1}{c|}{1} \\
         \cline{2-9}
         & $P_=$ & $P_\gamma$ & $P_\beta$ & $P_\alpha$& $P_\alpha$ & $P_\beta$ & $P_\gamma$ &  $P_=$ \\
    \end{tabular}
    \vspace{3ex}
    \caption{
        Illustration of $(P_\alpha, P_\beta, P_\gamma, P_=)$
        for three distinct nonzero vectors $\vec{\alpha}, \vec{\beta}, \vec{\gamma} \in \bbF_2^n$.
    }
    \label{tab:alpha-beta-gamma}
\end{minipage}%
\hfill%
\begin{minipage}{0.45\hsize}%
    \centering
    \includegraphics[width=0.9\linewidth]{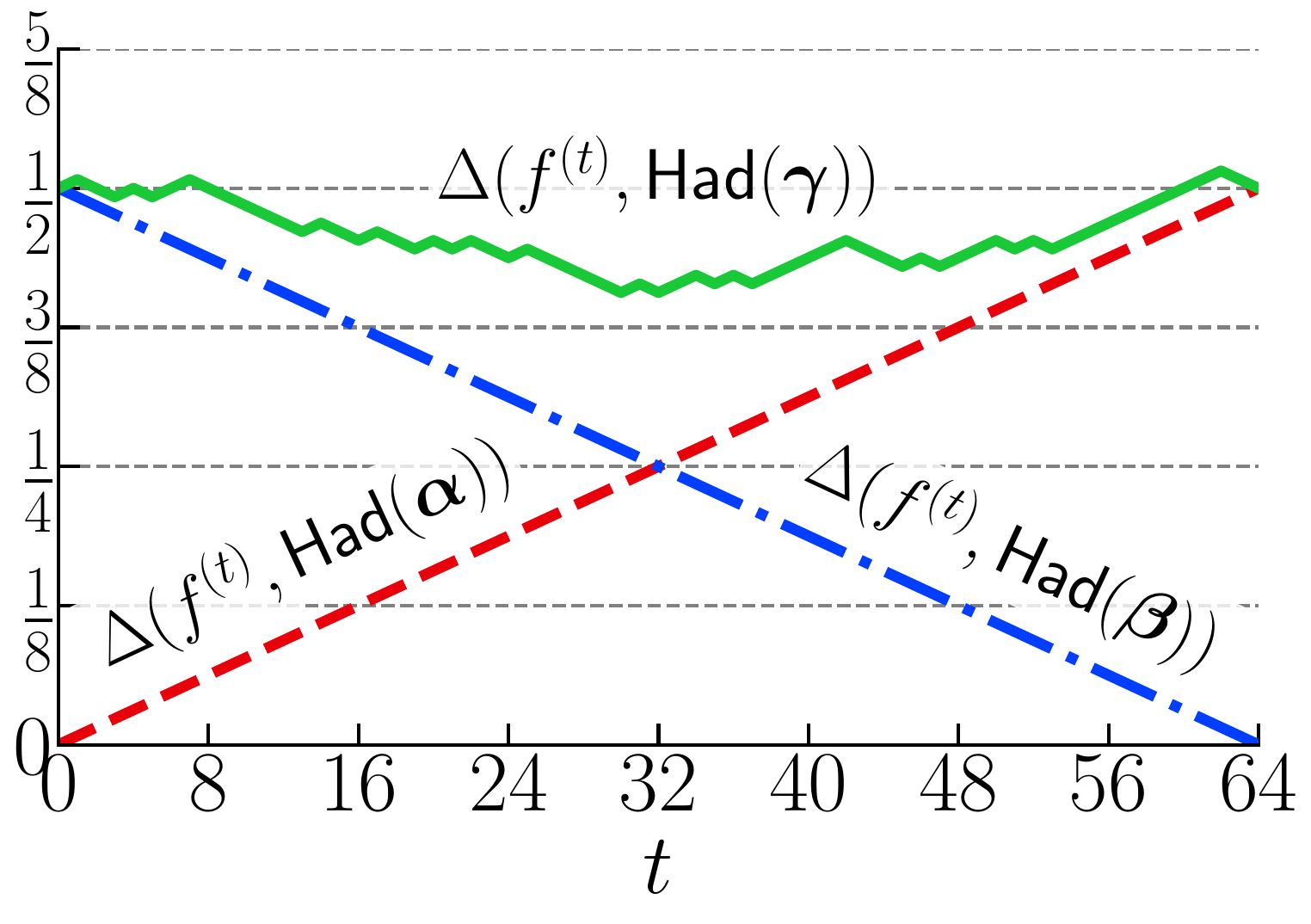}
    \vspace{-2ex}
    \caption{
        Plot of the distance from $f^{(\ttt)}$
        to $\Had(\vec{\alpha})$, $\Had(\vec{\beta})$, and $\Had(\vec{\gamma})$
        for a random reconfiguration $\Pi$ from $\Had(\vec{\alpha})$ to $\Had(\vec{\beta})$
        described in the proof of \cref{lem:Hadmard-reconf}.
    }
    \label{fig:Hadmard-reconf}
\end{minipage}%
\hfill\null%
\end{figure}

\subsection{Robustization}
\label{subsec:main:robust}

Subsequently, we advance to \emph{robustization} of \prb{Maxmin BCSP Reconfiguration},
relying on the reconfigurability of Hadamard codes.
For a system of Boolean circuits $\scrC$ and
its two satisfying truth assignments $\sigma^\ini, \sigma^\tar \colon \bbF_2^N \to \bbF_2$,
\prb{Circuit SAT Reconfiguration} requires to 
decide the existence of a reconfiguration sequence
from $\sigma^\ini$ to $\sigma^\tar$ over $\bbF_2^{\bbF_2^N}$ consisting only of
satisfying truth assignments to $\scrC$.

\begin{lemma}[Robustization]
\label{lem:robust}
There exists a polynomial-time algorithm that takes
an instance $(G,\psi^\ini,\psi^\tar)$ of \prb{Maxmin BCSP$_W$ Reconfiguration}
with alphabet size $W \in \bbN$, where
$\psi^\ini$ and $\psi^\tar$ satisfy $G$, and
then produces an instance $(\scrC, \sigma^\ini, \sigma^\tar)$ of \prb{Circuit SAT Reconfiguration}, where
$\scrC = (C_e)_{e \in E}$ is a system of circuits and
$\sigma^\ini$ and $\sigma^\tar$ satisfy $\scrC$, such that the following hold\textup{:}
\begin{itemize}
    \item \textup{(}Perfect completeness\textup{)} If $\val_G(\psi^\ini \reco \psi^\tar) = 1$,
    there exists a reconfiguration sequence from $\sigma^\ini$ to $\sigma^\tar$ made up of
    satisfying truth assignments to $\scrC$.
    \item \textup{(}Soundness\textup{)} If $\val_G(\psi^\ini \reco \psi^\tar) < 1-\epsilon$,
    any reconfiguration sequence from $\sigma^\ini$ to $\sigma^\tar$
    includes assignment $\sigma^\circ$
    such that
    for more than $\epsilon$-fraction of edges $e$ of $G$,
    $\sigma^\circ|_{\llbracket e \rrbracket}$ is $\frac{\delta_0}{8}$-far from
    any satisfying truth assignment to $C_e$,
    where $\delta_0 = \frac{1}{400}$ as in \cref{lem:Hadmard-reconf}.
\end{itemize}
\end{lemma}

\paragraph{Reduction.}
Our polynomial-time robustization of \prb{Maxmin BCSP$_W$ Reconfiguration}
into \prb{Circuit SAT Reconfiguration} is described as follows.
Let $(G,\psi^\ini,\psi^\tar)$ be an instance
of \prb{Maxmin BCSP$_W$ Reconfiguration}, where
$G = (V,E,\Sigma,\Pi)$ is a binary constraint graph, and
$\psi^\ini$ and $\psi^\tar$ satisfy $G$.
Without loss of generality, we can assume that
$W = |\Sigma| = 2^n$ for some integer $n \geq 9$,\footnote{
Otherwise, we can augment $\Sigma$ by padding so that $|\Sigma| \geq 2^9$.
}
and we can identify $\bbF_2^n$ with $\Sigma$.

Consider replacing binary constraints of $G$ by a system of circuits.
We first specify a \emph{truth assignment} to the entire circuit system
by a function $\sigma \colon \bbF_2^n \times V \to \bbF_2$,
which can be thought of as a concatenation of functions $\sigma_v \colon \bbF_2^n \to \bbF_2$
associated with each vertex $v \in V$.
For vertex $v \in V$,
let $\llbracket v \rrbracket$ denote the set of $2^n$ Boolean variables associated with $v$, and
for edge $e = (v,w) \in E$,
let $\llbracket e \rrbracket \triangleq \llbracket v \rrbracket \uplus \llbracket w \rrbracket$.\footnote{
Similar notations are used in \cite{dinur2007pcp}.
}
By this representation,
we can identify $\bbF_2^n \times V$ with $\biguplus_{v \in V} \llbracket v \rrbracket$.
In particular,
for edge $e=(v,w) \in E$,
$\sigma|_{\llbracket e \rrbracket}$
is equal to 
$\sigma|_{\llbracket v \rrbracket} \circ \sigma|_{\llbracket w \rrbracket}$.
For each edge $e=(v,w)$ of $G$ and its constraint $\pi_e$,
we define a circuit
$C_e \colon (\llbracket v \rrbracket \to \bbF_2) \times (\llbracket w \rrbracket \to \bbF_2) \to \bbF_2 $
(or equivalently, $C_e \colon \bbF_2^{\llbracket v \rrbracket} \times \bbF_2^{\llbracket w \rrbracket} \to \bbF_2$)
that depends \emph{only} on
$\sigma|_{\llbracket e \rrbracket} = \sigma|_{\llbracket v \rrbracket} \circ \sigma|_{\llbracket w \rrbracket}$
such that 
$C_e(\sigma|_{\llbracket v \rrbracket} \circ \sigma|_{\llbracket w \rrbracket}) = 1$ if and only if
\begin{align}
\begin{aligned}
    & \Delta(\sigma|_{\llbracket v \rrbracket}, \Had(\cdot)) \leq \frac{1}{4} \text{ and }
    \Delta(\sigma|_{\llbracket w \rrbracket}, \Had(\cdot)) \leq \frac{1}{4}, \\
    & \forall \alpha, \beta \in \Sigma,
    \Delta(\sigma|_{\llbracket v \rrbracket}, \Had(\alpha)) \leq \frac{1}{4} + \frac{\delta_0}{2} \text{ and }
    \Delta(\sigma|_{\llbracket w \rrbracket}, \Had(\beta)) \leq \frac{1}{4} + \frac{\delta_0}{2}
    \implies (\alpha, \beta) \in \pi_e,
\end{aligned}
\end{align}
where $\delta_0 = \frac{1}{400}$ as in \cref{lem:Hadmard-reconf}.
Note that each $C_e$ has constant size and can be constructed in constant time since $n = \bigO(1)$.
Consequently, we obtain
a system of circuits, denoted
$\scrC = (C_e)_{e \in E}$.
Given a satisfying assignment $\psi \colon V \to \Sigma$ for $G$,
we can construct a satisfying truth assignment $\sigma \colon \bbF_2^n \times V \to \bbF_2$
such that
$\sigma|_{\llbracket v \rrbracket} \triangleq \Had(\psi(v))$ for all $v \in V$.
Constructing $\sigma^\ini$ from $\psi^\ini$ and
$\sigma^\tar$ from $\psi^\tar$ according to this procedure,
we obtain an instance $(\scrC, \sigma^\ini, \sigma^\tar)$
of \prb{Circuit SAT Reconfiguration}.
Observe that the above reduction completes in polynomial time.

\begin{proof}[Proof of \cref{lem:robust}]
We first prove the perfect completeness.
It suffices to consider the case that 
$\psi^\ini$ and $\psi^\tar$ differ in exactly one vertex, say, $v^\star \in V$.
Using \cref{lem:Hadmard-reconf}, we obtain
a reconfiguration sequence
$( f^{(1)}, \ldots, f^{(\TTT)} )$
from $\Had(\psi^\ini(v))$ to $\Had(\psi^\tar(v))$.
Construct then a reconfiguration sequence
$\seqsigma = ( \sigma^{(1)}, \ldots, \sigma^{(\TTT)} )$
from $\sigma^\ini$ to $\sigma^\tar$
such that for all $\ttt$,
$\sigma^{(\ttt)}|_{\llbracket w \rrbracket} \triangleq \sigma^\ini|_{\llbracket w \rrbracket} = \sigma^\tar|_{\llbracket w \rrbracket}$
for all $w \neq v^\star$, and
$\sigma^{(\ttt)}|_{\llbracket v^\star \rrbracket} \triangleq f^{(\ttt)}$.
For each edge $e = (v^\star,w)$ of $G$,
any intermediate function $\sigma^{(\ttt)}$ of $\seqsigma$ satisfies the following:
\begin{itemize}
    \item By \cref{lem:Hadmard-reconf},
    $\sigma^{(\ttt)}|_{\llbracket v^\star \rrbracket}$ is
    $\frac{1}{4}$-close to $\Had(\psi^\ini(v))$ or $\Had(\psi^\tar(v))$, but 
    $\left(\frac{1}{4}+\delta_0\right)$-far from $\Had(\gamma)$
    for every $\gamma \notin \{\psi^\ini(v), \psi^\tar(v)\}$.
    \item $\sigma^{(\ttt)}|_{\llbracket w \rrbracket}$ is
    equal to $\Had(\psi^\ini(w)) = \Had(\psi^\tar(w))$; i.e.,
    it is $\left(\frac{1}{2}-o(1)\right)$-far from $\Had(\gamma)$
    for every $\gamma \notin \{\psi^\ini(w), \psi^\tar(w)\}$.
\end{itemize}
Since
$\{\psi^\ini(v^\star), \psi^\tar(v^\star)\} \times \{\psi^\ini(w), \psi^\tar(w)\}
= \{(\psi^\ini(v^\star), \psi^\ini(w)), (\psi^\tar(v^\star), \psi^\tar(w))\}
\subseteq \pi_e$,
it turns out that
$\sigma^{(\ttt)}|_{\llbracket e \rrbracket}$ satisfies $C_e$, and thus
every $\sigma^{(\ttt)}$ in $\seqsigma$ satisfies $\scrC$ entirely.

We then prove the soundness.
Suppose
$\val_G(\psi^\ini \reco \psi^\tar) < 1-\epsilon$ and
we are given a reconfiguration sequence
$\seqsigma = ( \sigma^{(1)}, \ldots, \sigma^{(\TTT)} )$
from $\sigma^\ini$ to $\sigma^\tar$.
Construct then a reconfiguration sequence
$\seqpsi = ( \psi^{(1)}, \ldots, \psi^{(\TTT)} )$
from $\psi^\ini$ to $\psi^\tar$ such that
$\psi^{(\ttt)}(v)$ is defined as a value of $\Sigma$ whose Hadamard codeword is closest to
$\sigma^{(\ttt)}|_{\llbracket v \rrbracket}$; namely,\footnote{
Ties are broken according to any prefixed order of $\Sigma$.
}
\begin{align}
    \psi^{(\ttt)}(v) \triangleq
    \argmin_{\alpha \in \Sigma} \Delta(\sigma^{(\ttt)}|_{\llbracket v \rrbracket}, \Had(\alpha)).
\end{align}
Since $\seqpsi$ is a valid reconfiguration sequence,
there exists some $\psi^{(\ttt)}$
that violates more than $\epsilon \cdot |E|$ edges.

Hereafter, we denote
$\psi \triangleq \psi^{(\ttt)}$ and
$\sigma \triangleq \sigma^{(\ttt)}$
for notational simplicity.
Suppose $\psi$ violates edge $e = (v,w)$; i.e.,
    $(\psi(v), \psi(w)) \notin \pi_e$.
We would like to show that
$\sigma|_{\llbracket e \rrbracket}$ is
$\frac{\delta_0}{8}$-far from any satisfying truth assignment to $C_e$.
Let $f,g \colon \bbF_2^n \to \bbF_2$ by a satisfying truth assignment to $C_e$.
In particular, there exists a pair $(\alpha^\star, \beta^\star) \in \pi_e$ such that
$\Delta(f, \Had(\alpha^\star)) \leq \frac{1}{4}$ and
$\Delta(g, \Had(\beta^\star)) \leq \frac{1}{4}$.
Observe now that 
``$f$ is $\left(\frac{1}{4} + \frac{\delta_0}{2}\right)$-far from $\Had(\psi(v))$'' or
``$g$ is $\left(\frac{1}{4} + \frac{\delta_0}{2}\right)$-far from $\Had(\psi(w))$''
because otherwise, $C_e(f \circ g) = 0$.

Suppose first $\Delta(f, \Had(\psi(v))) > \frac{1}{4} + \frac{\delta_0}{2}$,
implying that $\alpha^\star \neq \psi(v)$.
Putting together, we have the following three inequalities in hand:
\begin{align}
    \Delta(f, \Had(\alpha^\star)) & \leq \frac{1}{4}
        & \text{by assumption}, \\
    \Delta(f, \Had(\psi(v))) & > \frac{1}{4} + \frac{\delta_0}{2}
        & \text{by assumption}, \\
%    \Delta(\sigma(v), \Had(\psi(v))) & \leq \frac{1}{4} + \frac{\delta_0}{2} by assumption \\
    \Delta(\sigma|_{\llbracket v \rrbracket}, \Had(\psi(v))) & \leq \Delta(\sigma|_{\llbracket v \rrbracket}, \Had(\alpha^\star))
        & \text{by construction of }\sigma|_{\llbracket v \rrbracket}.
\end{align}
Simple calculation using the triangle inequality derives
\begin{align}
    &
    \begin{aligned}
    \Delta(f, \Had(\psi(v)))
    & \leq \Delta(f, \sigma|_{\llbracket v \rrbracket}) + \Delta(\sigma|_{\llbracket v \rrbracket}, \Had(\psi(v))) \\
    & \leq \Delta(f, \sigma|_{\llbracket v \rrbracket}) + \Delta(\sigma|_{\llbracket v \rrbracket}, \Had(\alpha^\star)) \\
    & \leq \Delta(f, \sigma|_{\llbracket v \rrbracket}) + \Delta(\sigma|_{\llbracket v \rrbracket}, f) + \Delta(f, \Had(\alpha^\star)) \\
    & = 2 \cdot \Delta(\sigma|_{\llbracket v \rrbracket}, f) + \Delta(f, \Had(\alpha^\star))
    \end{aligned} \\
    &
    \implies
    2 \cdot \Delta(\sigma|_{\llbracket v \rrbracket}, f)
    \geq 
    \underbrace{\Delta(f, \Had(\psi(v)))}_{> \frac{1}{4} + \frac{\delta_0}{2}}
    - \underbrace{\Delta(f, \Had(\alpha^\star))}_{\leq \frac{1}{4}} \\
    &
    \implies
    \Delta(\sigma|_{\llbracket v \rrbracket}, f)
    > \frac{\delta_0}{4}.
\end{align}
Consequently, $\sigma|_{\llbracket e \rrbracket}$ should be $\frac{\delta_0}{8}$-far from
$f \circ g$.

Suppose next $\Delta(g, \Had(\psi(w))) > \frac{1}{4} + \frac{\delta_0}{2}$,
implying that $\beta^\star \neq \psi(w)$.
Similarly to the first case,
we can show that
$\Delta(\sigma|_{\llbracket w \rrbracket}, g) > \frac{\delta_0}{4}$,
deriving that
$\sigma|_{\llbracket e \rrbracket}$ is $\frac{\delta_0}{8}$-far from $f \circ g$.
This completes the proof of the soundness.
\end{proof}

\cref{eg:robust:detail} explains
why the reconfigurability of Hadamard codes is needed,
by using a slightly different definition of circuits that fails robustization.

\begin{example}
\label{eg:robust:detail}
For edge $e = (v,w)$ of $G$, define a binary constraint
$\pi_e \triangleq \{(\alpha_1, \beta_1), (\alpha_2, \beta_2)\} \subset \bbF_2^n \times \bbF_2^n$.
Construct a circuit
$\tilde{C}_e \colon (\llbracket v\rrbracket \to \bbF_2) \times (\llbracket w\rrbracket \to \bbF_2) \to \bbF_2$
such that
$\tilde{C}_e(\sigma|_{\llbracket v\rrbracket} \circ \sigma|_{\llbracket w \rrbracket}) = 1$
if and only if
\begin{align}
\begin{aligned}
    & \Delta(\sigma|_{\llbracket v \rrbracket}, \Had(\cdot)) \leq \frac{1}{4}
    \text{ and }
    \Delta(\sigma|_{\llbracket w \rrbracket}, \Had(\cdot)) \leq \frac{1}{4}, \\
    & \forall \alpha, \beta \in \Sigma,
    \Delta(\sigma|_{\llbracket v \rrbracket}, \Had(\alpha)) \leq \frac{1}{4} \text{ and }
    \Delta(\sigma|_{\llbracket w \rrbracket}, \Had(\beta)) \leq \frac{1}{4}
    \implies
    (\alpha, \beta) \in \pi_e.
\end{aligned}
\end{align}
Note that reconfiguring from
$(\alpha_1, \beta_1)$ to
$(\alpha_2, \beta_2)$ over $\Sigma \times \Sigma$
(not $\bbF_2^n \times \bbF_2^n$)
must break $\pi_e$ (at some point).
Analogously,
we might expect that any reconfiguration sequence from 
$\Had(\alpha_1) \circ \Had(\beta_1)$ to
$\Had(\alpha_2) \circ \Had(\beta_2)$ over $\bbF_2^{\llbracket v \rrbracket} \times \bbF_2^{\llbracket w \rrbracket}$
includes a function that is $\Theta(1)$-far from any 
satisfying truth assignment to $\tilde{C}_e$.
Consider now the following reconfiguration:

\begin{itembox}[l]{\textbf{Reconfiguration $\Pi$
from $\Had(\alpha_1) \circ \Had(\beta_1)$ to $\Had(\alpha_2) \circ \Had(\beta_2)$.}}
\begin{algorithmic}[1]
\State $f \triangleq$ a function $\frac{1}{4}$-close to both $\Had(\alpha_1)$ and $\Had(\alpha_2)$.
\State $g \triangleq$ a function $\frac{1}{4}$-close to both $\Had(\beta_1)$ and $\Had(\beta_2)$.
%\LComment{start from $\Had(\alpha_1) \circ \Had(\beta_1)$}.
\State change $\Had(\alpha_1)$ to $f$ one by one.
\State change $\Had(\beta_1)$ to $g$ one by one.
\LComment{obtain $f \circ g$.}
\State change $f$ to $\Had(\alpha_2)$ one by one.
\State change $g$ to $\Had(\beta_2)$ one by one.
%\LComment{end at $\Had(\alpha_2) \circ \Had(\beta_2)$.}
\end{algorithmic}
\end{itembox}
Changing particular two bits of $f \circ g$,
we obtain
$f^\star \circ g^\star$, which is 
$\left(\frac{1}{4}-\frac{1}{2^n}\right)$-close to $\Had(\alpha_1) \circ \Had(\beta_1)$,
implying $\tilde{C}_e(f^\star \circ g^\star) = 1$.
Thus,
$f \circ g$ is $\frac{1}{2^n}$-close to some satisfying truth assignment to $\tilde{C}_e$.
Similarly, every intermediate function of $\Pi$ is
$\frac{1}{2^n}$-close to some satisfying truth assignment to $\tilde{C}_e$.
\lipicsEnd
\end{example}

\subsection{Composition of Assignment Testers}
\label{subsec:main:composition}

We are now ready to compose an assignment tester into \prb{Circuit SAT Reconfiguration}
to accomplish alphabet reduction of \prb{Maxmin BCSP Reconfiguration}.
Here, we recapitulate \emph{assignment testers} \cite{dinur2007pcp,dinur2006assignment},
a.k.a.~\emph{PCPs of proximity} \cite{bensasson2006robust},
and refer to an explicit construction due to \citet{dinur2007pcp,odonnell2014analysis}.\footnote{
Note that an assignment tester of \citet[Theorem~7.16]{odonnell2014analysis}
takes the form of verifiers,
which can be represented as a binary constraint graph by
a standard reduction from probabilistically checkable proofs to two-prover games,
e.g., \cite{fortnow1994power,radhakrishnan2007dinur}.
}

\begin{definition}[\cite{dinur2006assignment,bensasson2006robust}]
\label{def:AT}
An \emph{assignment tester} over
alphabet $\Sigma_0 \supset \bbF_2$ with
\emph{rejection rate} $\rho \in (0,1)$
is an algorithm $\calP$ that takes
a circuit $\Phi \colon \bbF_2^X \to \bbF_2$
over Boolean variables $X$ as input, and
produces
a binary constraint graph $G = (V=X \uplus Y, E, \Sigma_0, \Pi)$
over $X$ and auxiliary variables $Y$ such that
the following hold
for any truth assignment $\sigma \colon X \to \bbF_2$ for $\Phi$:
\begin{itemize}
    \item (Perfect completeness)
    If $\sigma$ satisfies $\Phi$,
    there exists an assignment
    $\tau \colon Y \to \Sigma_0$ such that
    $\val_G(\sigma \circ \tau) = 1$.
    \item (Soundness)
    If $\sigma$ is $\delta$-far from any satisfying truth assignment to $\Phi$,
    for every assignment 
    $\tau \colon Y \to \Sigma_0$,
    $\val_G(\sigma \circ \tau) < 1- \rho \cdot \delta$.
    \lipicsEnd
\end{itemize}
\end{definition}

\begin{theorem}[\protect{\cite[Theorem~5.1]{dinur2007pcp} and \cite[Theorem~7.16]{odonnell2014analysis}}]
\label{thm:AT}
There exists an explicit construction of an assignment tester $\calP$
with alphabet $\Sigma_0 = \bbF_2^3$ and rejection rate $\rho \triangleq \frac{1}{10{,}000}$.
\end{theorem}

\begin{proposition}[Composition]
\label{prp:composition}
There exist
universal constants $\tilde{W}_0 \triangleq 8$ and $\tilde{\kappa} \triangleq \frac{\delta_0^2 \rho^2}{64} \in (0,1)$, and
a polynomial-time algorithm that takes
an instance
$(G,\psi^\ini,\psi^\tar)$
of \prb{Maxmin BCSP$_W$ Reconfiguration}
with alphabet size $W \in \bbN$, where
$\psi^\ini$ and $\psi^\tar$ satisfy $G$, and
then produces an instance
$(G',\psi'^\ini,\psi'^\tar)$
of \prb{Maxmin $4$-CSP$_{\tilde{W}_0}$ Reconfiguration}
with alphabet size $\tilde{W}_0$, where
$\psi'^\ini$ and $\psi'^\tar$ satisfy $G'$,
such that the following hold\textup{:}
\begin{itemize}
    \item \textup{(}Perfect completeness\textup{)}
    If $\val_G(\psi^\ini \reco \psi^\tar) = 1$, then
    $\val_{G'}(\psi'^\ini \reco \psi'^\tar) = 1$.
    \item \textup{(}Soundness\textup{)} 
    If $\val_G(\psi^\ini \reco \psi^\tar) < 1-\epsilon$, then
    $\val_{G'}(\psi'^\ini \reco \psi'^\tar) < 1- \tilde{\kappa} \cdot \epsilon$.
\end{itemize}
\end{proposition}

\paragraph{Reduction.}
We now describe a polynomial-time reduction from
\prb{Circuit SAT Reconfiguration} introduced in the previous subsection
to
\prb{Maxmin $4$-CSP$_8$ Reconfiguration}.
Let $(\scrC, \sigma^\ini, \sigma^\tar)$ be an instance of
\prb{Circuit SAT Reconfiguration}
obtained by applying \cref{lem:robust} to an instance $(G, \psi^\ini, \psi^\ini)$ of
\prb{Maxmin BCSP$_W$ Reconfiguration}.
Here, $\scrC = (C_e)_{e \in E}$ is a system of circuits
over Boolean variables $\bbF_2^n \times V$,
associated with underlying graph $(V,E)$, and
$\sigma^\ini$ and $\sigma^\tar$ entirely satisfy $\scrC$.

Running the assignment tester $\calP$ of \cref{thm:AT} on
each circuit $C_e \colon \bbF_2^{\llbracket e \rrbracket} \to \bbF_2$ for edge $e \in E$
produces a binary constraint graph
$G_e = (V_e = \llbracket e \rrbracket \uplus Y_e, E_e, \Sigma_0, \tilde{\Pi}_e = (\tilde{\pi}_{\tilde{e}})_{\tilde{e} \in E_e})$, where
$Y_e$ is the set of auxiliary variables and $|\Sigma_0| = 8$.
Create a pair of copies of $G_e$ ``sharing'' $\llbracket e \rrbracket$,
denoted $G_e^1$ and $G_e^2$; namely,
\begin{align}
    G_e^1 & \triangleq (V_e^1 = \llbracket e \rrbracket \uplus Y_e^1, E_e^1, \Sigma_0, \tilde{\Pi}_e^1), \\
    G_e^2 & \triangleq (V_e^2 = \llbracket e \rrbracket \uplus Y_e^2, E_e^2, \Sigma_0, \tilde{\Pi}_e^2).
\end{align}
We then ``superimpose'' $G_e^1$ and $G_e^2$
to obtain a $4$-ary constraint graph
$G'_e = (V'_e, E'_e, \Sigma_0, \Pi'_e = (\pi'_{(\tilde{e}_1,\tilde{e}_2)})_{(\tilde{e}_1,\tilde{e}_2) \in E_e})$, where
\begin{align}
\begin{aligned}
    V'_e & \triangleq \llbracket e \rrbracket \uplus Y_e^1 \uplus Y_e^2,
    \text{ and } E'_e \triangleq E_e^1 \times E_e^2, \\
    \pi'_{(\tilde{e}_1,\tilde{e}_2)} & \triangleq
    \tilde{\pi}_{\tilde{e}_1} \times \tilde{\pi}_{\tilde{e}_2}
    = \Bigl\{
        (\alpha_1, \beta_1, \alpha_2, \beta_2) \in \Sigma^4 \Bigm|
        (\alpha_1, \beta_1) \in \tilde{\pi}_{\tilde{e}_1} \vee
        (\alpha_2, \beta_2) \in \tilde{\pi}_{\tilde{e}_2}
    \Bigr\} \\
    & \;\;\;\; \text{ for all } (\tilde{e}_1,\tilde{e}_2) \in E_e^1 \times E_e^2.
\end{aligned}
\end{align}
Note that
each pair of edges from $E_e^1$ and $E_e^2$ forms
a hyperedge of $G'_e$,
which would be satisfied if so is either of the two edges.
We can safely assume that $E'_e$ has the same size for all $e \in E$.

Finally, the new $4$-ary constraint graph $G' = (V', E', \Sigma_0, \Pi')$ is defined as
\begin{align}
\begin{aligned}
    V' & \triangleq \bigcup_{e \in E} V'_e
        = \left (\biguplus_{v \in V} \llbracket v \rrbracket \right)
        \uplus \left(\biguplus_{e \in E} Y_e^1 \uplus Y_e^2\right), \\
    E' & \triangleq \biguplus_{e \in E} E'_e
    \text{ and }
    \Pi' \triangleq \biguplus_{e \in E} \Pi'_e.
\end{aligned}
\end{align}
For any satisfying truth assignment
$\sigma \colon \biguplus_{v \in V} \llbracket v \rrbracket \to \bbF_2$ of $\scrC$,
consider an assignment $\psi' \colon V' \to \Sigma_0$ such that
$\psi'|_{\llbracket v \rrbracket} \triangleq \sigma|_{\llbracket v \rrbracket}$
for all $v \in V$ and
$\psi'|_{Y_e^1} = \psi'|_{Y_e^2} = \tau_e$
for all $e \in E$,
where $\tau_e \colon Y_e \to \Sigma_0$ is an assignment to auxiliary variables $Y_e$
such that
$\sigma|_{\llbracket e \rrbracket} \circ \tau_e$ satisfies $G'_e$,
whose existence is guaranteed by \cref{def:AT}.
Observe easily that $\psi'$ satisfies $G'$.
Constructing $\psi'^\ini$ from $\sigma^\ini$
and $\psi'^\tar$ from $\sigma^\tar$ according to this procedure,
we obtain an instance $(G',\psi'^\ini, \psi'^\tar)$ of
\prb{Maxmin $4$-CSP$_8$ Reconfiguration},
completing the reduction.

\begin{proof}[Proof of \cref{prp:composition}]
Recall that
$(G,\psi^\ini,\psi^\tar)$ is an instance of \prb{Maxmin BCSP$_W$ Reconfiguration},
$(\scrC,\sigma^\ini,\sigma^\tar)$ is an instance of \prb{Circuit SAT Reconfiguration}
obtained by applying \cref{lem:robust}, and
$(G',\psi'^\ini,\psi'^\tar)$ is an instance of \prb{Maxmin $4$-CSP$_8$ Reconfiguration}
obtained by composing the assignment tester \cite{dinur2007pcp} as described above.

We first prove the perfect completeness.
By \cref{lem:robust},
it suffices to consider the case that
$\sigma^\ini$ and $\sigma^\tar$ differ in exactly one variable,
say, $ (\vec{x}, v^\star) \in \bbF_2^n \times V$.
Consider a reconfiguration sequence $\seqpsi'$
from $\psi'^\ini$ to $\psi'^\tar$
obtained by the following procedure:
\begin{itembox}[l]{\textbf{Reconfiguration $\seqpsi'$ from $\psi'^\ini$ to $\psi'^\tar$.}}
\begin{algorithmic}[1]
    \ForAll{edge $e = (v^\star,w) \in E$}
        \State let $\tau_e^\tar \colon Y_e \to \Sigma_0$
        be assignment such that
        $\sigma^\tar|_{\llbracket e \rrbracket} \circ \tau_e^\tar$ satisfies $G_e$.
        \State change the entries on $Y_e^1$ to $\tau_e^\tar$ one by one.
    \EndFor
    \item flip \nth{$\vec{x}$} entry of $\llbracket v^\star \rrbracket$.
    \ForAll{edge $e = (v^\star,w) \in E$}
        \State change the entries on $Y_e^2$ to $\tau_e^\tar$ one by one.
    \EndFor
\end{algorithmic}
\end{itembox}
Observe easily that for any edge $e = (v^\star,w) \in E$,
either of $G_e^1$ or $G_e^2$ is entirely satisfied
by any intermediate assignment, implying that
$\val_{G'}(\seqpsi') = 1$,
as desired.

We then prove the soundness.
Suppose we are given a reconfiguration sequence
$\seqpsi' = ( \psi'^{(1)}, \ldots, \psi'^{(\TTT)} )$
from $\psi'^\ini$ to $\psi'^\tar$ such that
$\val_{G'}(\seqpsi') = \val_{G'}(\psi'^\ini \reco \psi'^\tar)$.
Consider a reconfiguration sequence 
$\seqsigma = ( \sigma^{(1)}, \ldots, \sigma^{(\TTT)} )$ such that
$\sigma^{(\ttt)} \triangleq \psi'^{(\ttt)}|_{\biguplus_{v \in V}\llbracket v \rrbracket}$
for all $\ttt$.
Since $\seqsigma$ is
a valid reconfiguration sequence from $\sigma^\ini$ to $\sigma^\tar$,
by \cref{lem:robust},
there exists some $\sigma^{(\ttt)}$ such that
for more than $\epsilon$-fraction of edges $e$ of $G$,
$\sigma^{(\ttt)}|_{\llbracket e \rrbracket} = \psi'^{(\ttt)}|_{\llbracket e \rrbracket}$
is $\frac{\delta_0}{8}$-far from any satisfying truth assignment to $C_e$.
Let $F \subset E$ be the set of such edges of $G$; note that $|F| \geq \epsilon |E|$.
By \cref{thm:AT},
$\psi'^{(\ttt)}$ violates more than $\frac{\delta_0 \rho}{8}$-fraction of edges of each $G_e^1$ and $G_e^2$
for any $e \in F$.
Since $\psi'^{(\ttt)}$ violates
hyperedge $(\tilde{e}_1, \tilde{e}_2) \in E_e^1 \times E_e^2$
if and only if
it violates
$\tilde{e}_1 \in E_e^1$ with respect to $\tilde{\Pi}^1_e$ and
$\tilde{e}_2 \in E_e^2$ with respect to $\tilde{\Pi}^2_e$
\emph{simultaneously},
there are
more than $\left(\frac{\delta_0 \rho}{8}\right)^2 $-fraction of hyperedges of $G'_e$
that are violated by
$\psi'^{(\ttt)}$; i.e., $1-\val_{G_e}(\psi'^{(\ttt)}) > \frac{\delta_0^2 \rho^2}{64}$.
Consequently, we derive
\begin{align}
\begin{aligned}
    1-\val_{G'}(\seqpsi')
    & \geq 1-\val_{G'}(\psi'^{(\ttt)}) \\
    & = \frac{1}{|E|} \sum_{e \in E} \Bigl( 1-\val_{G'_e}(\psi'^{(\ttt)}) \Bigr)
        & \text{(since every } E_e \text{ has the same size)} \\
    & \geq \frac{1}{|E|} \sum_{e \in F} \Bigl( 1-\val_{G'_e}(\psi'^{(\ttt)}) \Bigr) \\
    & > \frac{|F|}{|E|} \frac{\delta_0^2 \rho^2}{64}
    > \epsilon \cdot \underbrace{\frac{\delta_0^2 \rho^2}{64}}_{=\tilde{\kappa}},
\end{aligned}
\end{align}
implying that
$\val_{G'}(\psi'^\ini \reco \psi'^\tar) = \val_{G'}(\seqpsi') < 1 - \tilde{\kappa} \cdot \epsilon$,
as desired.
\end{proof}

\begin{proof}[Proof of \cref{thm:ABC-reduct}]
Our construction of alphabet reduction for \prb{Maxmin BCSP Reconfiguration} follows from
\cref{lem:robust,prp:composition} and
a gap-preserving reduction \cite[Lemma 5.4]{ohsaka2024gap}
(which is in fact approximation-preserving) from 
\prb{Gap$_{1,1-\epsilon}$ $4$-CSP$_{\tilde{W}_0}$ Reconfiguration} to
\prb{Gap$_{1,1-\frac{\epsilon}{4}}$ BCSP$_{W_0}$ Reconfiguration},
where $W_0 = \left(\frac{\tilde{W}_0(\tilde{W}_0+1)}{2}\right)^4 = 36^4$.
The value of $\kappa$ in \cref{thm:ABC-reduct} should be
$\frac{\tilde{\kappa}}{4} = \frac{\delta_0^2 \rho^2}{256} = \frac{1}{256 \cdot 400^2 \cdot 10{,}000^2} = \frac{1}{8{,}000^4}$.
\end{proof}

\section{Conclusions}
\label{sec:conclusions}
We presented \citeauthor{dinur2007pcp}'s style alphabet reduction \cite{dinur2007pcp}
for \prb{Maxmin Binary CSP Reconfiguration},
which now makes both the degree of inapproximability and alphabet size oblivious to 
the (arbitrarily small) gap parameter of RIH \cite{ohsaka2023gap}.
The main ingredient of its construction is
the \emph{reconfigurability of Hadamard codes},
which may be of independent interest and have further applications.
We leave some open questions:
\begin{itemize}
    \item \textbf{(Question 1)}
    Can we prove RIH \cite{ohsaka2023gap}
    by \citeauthor{dinur2007pcp}'s style gap amplification \cite{dinur2007pcp}?
    As discussed in \cref{subsec:intro:dinur},
    an approximation-preserving version for
    degree reduction and gap amplification of \prb{Maxmin Binary CSP Reconfiguration} \cite{ohsaka2023gap,ohsaka2024gap}
    seems mandatory.
    \item \textbf{(Question 2)}
    Can we derive more meaningful inapproximability factors?
    Alas, we acknowledge that
    the current inapproximability factor is so small as to be almost meaningless in practice.
    
    \item \textbf{(Question 3)}
    Given the reconfigurability of Hadamard codes (\cref{lem:Hadmard-reconf}),
    it is natural to ask that of other error-correcting codes:
    One may say that an error-correcting code $\enc$ is \emph{$(\delta,\mu)$-reconfigurable}
    if for any $\vec{\alpha} \neq \vec{\beta}$,
    there exists a reconfiguration sequence
    from $\enc(\vec{\alpha})$ to $\enc(\vec{\beta})$ such that
    every function in it is 
    \begin{itemize}
        \item $\delta$-close to either $\enc(\vec{\alpha})$ or $\enc(\vec{\beta})$, and
        \item $\left(\delta+\mu\right)$-far from $\enc(\vec{\gamma})$ for every $\vec{\gamma} \neq \vec{\alpha},\vec{\beta}$.
    \end{itemize}
    Is there any such reconfigurable error-correcting code?
    Also, is there any general composition scheme for probabilistically checkable reconfiguration proofs \cite{hirahara2024probabilistically}?
\end{itemize}

\paragraph{Acknowledgments.}
I wish to thank Shuichi Hirahara for helpful conversations, and
thank the anonymous referees for letting me know
a simple construction of an assignment tester due to 
\citet[Theorem~7.16]{odonnell2014analysis}.

\clearpage
\printbibliography

\end{document}